\documentclass[10pt, conference]{IEEEtran}

\usepackage{mystyle}
\usepackage{tabularx}

\usepackage{multicol}
\usepackage{supertabular}

\ifCLASSINFOpdf
\else
\fi

\begin{document}

\title{Structural Self-adaptation for Decentralized Pervasive Intelligence\vspace{-0.3cm}}

\author{\IEEEauthorblockN{Jovan Nikoli\'c, Evangelos Pournaras}
\IEEEauthorblockA{
ETH Zurich\\
Zurich, Switzerland\\
\{jnikolic,epournaras\}@ethz.ch\vspace{-0.3cm}}
}

\maketitle

\begin{abstract}
Communication structure plays a key role in the learning capability of decentralized systems. Structural self-adaptation, by means of self-organization, changes the order as well as the input information of the agents' collective decision-making. This paper studies the role of agents' repositioning on the same communication structure, i.e. a tree, as the means to expand the learning capacity in complex combinatorial optimization problems, for instance, load-balancing power demand to prevent blackouts or efficient utilization of bike sharing stations. The optimality of structural self-adaptations is rigorously studied by constructing a novel large-scale benchmark that consists of 4000 agents with synthetic and real-world data performing 4 million structural self-adaptations during which almost 320 billion learning messages are exchanged. Based on this benchmark dataset, 124 deterministic structural criteria, applied as learning meta-features, are systematically evaluated as well as two online structural self-adaptation strategies designed to expand learning capacity. Experimental evaluation identifies metrics that capture agents with influential information and their optimal positioning. Significant gain in learning performance is observed for the two strategies especially under low-performing initialization. Strikingly, the strategy that triggers structural self-adaptation in a more exploratory fashion is the most cost-effective. 
\end{abstract}

\IEEEpeerreviewmaketitle

\section{Introduction}\label{sec:introduction}

%

The rise of distributed pervasive intelligence in the Internet of Things provides new unprecedented means to perform decentralized optimization and learning over communication networks~\cite{Sadri2011}. Autonomous agents running on embedded devices can collaboratively solve complex optimization and learning problems without involvement of centralized third parties that do not scale, are single points of failure, require trust and privacy-sensitive data~\cite{pournaras2018}. The critical role that structure plays in conventional machine learning algorithms has been earlier underlined, for instance, the number of layers in neural networks or the dropout of neurons to prevent over-fitting~\cite{walczak1999,karsoliya2012}. However, little is known how communication structure influences decentralized learning in multi-agent systems. 

This paper fills this gap by studying \emph{fixed and dynamic structural self-adaptations} as the means of improving learning performance in challenging decentralized combinatorial optimization problems. Fixed adaptations concern deterministic criteria with which agents are repositioned within the same communication structure. This influences the order and the input information in agents' collective decision-making. Dynamic adaptations concern a deterministic or random repositioning of agents during runtime to explore higher-performing solutions and escape from suboptimal trapped solutions. A novel methodology is introduced to study optimality of decentralized collective learning under structural self-adaptations. It relies on empirical data from real-world pilot projects for load-balancing power demand and bike sharing stations. 

The \textbf{contributions of this paper} are outlined as follows: (i) A formal modeling approach of structural self-adaptation as a bijection of isomorphic graphs. (ii) A comparison of 124 structural self-adaptation criteria used as meta-features to improve offline or online learning performance. (iii) A modeling approach for designing online structural self-adaptations. (iv) Two customizable structural self-adaptation strategies to improve online learning performance. (v) A methodological approach to study the optimality of large-scale combinatorial optimization independent of data and application. (vi) An open benchmark dataset~\cite{Pournaras2019} of synthetic and real-world data for optimality evaluation. It contains 4 million performance profiles of structural self-adaptations generating almost 320 billion learning interactions among 4000 agents. (vii) Findings on the role of structural self-adaptations in learning aspects: optimality, application scenarios, network topology, self-adaptation parameters as well as computational and communication cost. 

This paper is outlined as follows: Section~\ref{sec:research-positioning} positions this work and Section~\ref{sec:related-work} outlines related work. Section~\ref{sec:structural-self-adaptation} formalizes fixed and dynamic structural self-adaptations. Section~\ref{sec:fixed-bijections} and~\ref{sec:dynamic-bijections} discuss each of them respectively. Two online structural self-adaptation strategies are designed in Section~\ref{sec:strategies}. The experimental methodology is outlined in Section~\ref{sec:methodology} and the experimental evaluation in Section~\ref{sec:evaluation}. Finally, Section~\ref{sec:conclusion} concludes this paper and outlines future work. 

\section{Research Positioning}\label{sec:research-positioning}

This paper studies collective decision-making in multi-agent systems, in which agents have a set of discrete options to choose from. These options are resource consumption or production \emph{plans} that are used for resource scheduling and allocation. For instance, a plan \plan can represent when a user charges its electric vehicle~\cite{Pournaras2017b}, the household energy demand over time, or the bike sharing stations from which a user picks up a bike or leaves one~\cite{pournaras2018}. In practice, plans are sequences of real values. Each agent \agent has multiple plans \planset{\agent} that model the flexibility of the agent, its alternative options. Each agent may have preferences over its plans measured by a \emph{local cost} assigned to each plan. For instance, the distance of a user from different bike sharing stations can measure the costs of plans. 

An agent's plan selection over the discrete planning options can satisfy \emph{local and global objectives}, which can often be orthogonal to each other. Meeting local objectives is about choosing the plan with the lowest local cost, while global objectives concern the minimization of a \emph{global cost} function with input the aggregate of all selected plans, i.e. element-wise summation. Agents can self-determine their preferences over the two objectives. On the one hand, linear cost functions can be minimized locally without coordinating the agents' plan selections. For instance, minimizing the total power demand in the Smart Grid is a result of locally choosing the plan with the lowest power demand. On the other hand, minimization of quadratic cost functions, such as the minimization of variance, is a challenging non-convex combinatorial optimization problem that is NP-hard with complexity $O(k^{\numagents})$, where $k$ is the number of plans per agent and $\numagents$ is the total number of agents~\cite{Rockafellar2000,Allen2016,pournaras2018}. Coordination between the agents' plan selections is required. The variance is used in this paper as a \emph{balancing indicator}, e.g. lowering power peaks to prevent blackouts~\cite{Thapa2017} or preserving a uniform number of bikes available at all bike sharing stations~\cite{pournaras2018}. Methods that parallelize computations are one approach to solve such complex computational problems, for instance, BnB-ADOPT~\cite{Yeoh2010}, NCBB~\cite{Chechetka2006} and DPOP~\cite{Petcu2005}. However, such methods require universal access to agents' information and therefore they are not designed for decentralized multi-agent systems preserving agents' privacy and autonomy. 

The alternative approach is to introduce a self-organzing communication structure that orchestrates in a cooperative and decentralized way searching of the combinatorial space~\cite{Ye2017}. This paper studies the role of such an agents' structure in performance. It focuses on a certain \emph{structural self-adaptation} that is \emph{the repositioning of the agents in a fixed topology as the means to explore the combinatorial solutions space}. Agents' positioning governs the input information from other agents based on which plan selections are made. In hierarchical structures such as trees, agents often interact in a bottom-up and top-down fashion~\cite{Diaconescu2018}. The agents' positioning governs the order of decision-making. A certain sequence of decision-making is a actually a coordination pattern: a traversal of the combinatorial solutions space. Different traversal strategies, i.e. criteria of agents' positioning, may be trapped in different suboptimal solutions with different performance status. 

It is fair to underline the challenge of determining causal relationships between performance and structure. Algorithmic artifacts introduce biases that are hard to distinguish from the role that the underlying structure plays. Given this challenge, this paper studies the decentralized collective learning system of I-EPOS\footnote{Available at
http://epos-net.org (last accessed: March 2019).}, the \emph{Iterative Economic Planning and Optimized Selections}~\cite{pournaras2018,Pournaras2019c}. The interactive agents of I-EPOS self-organize in tree communication structures to coordinate plan selections. The exact mechanics of this coordination is subject of earlier work~\cite{pournaras2018}. The structural self-adaptations studied in this paper are independent of I-EPOS that is used as a black box and a benchmark scenario given the following: (i) High efficiency\footnote{The cost-effectiveness of I-EPOS is characterized by the following~\cite{pournaras2018}: convergence to very few iterations, solutions of monotonously decreasing cost during convergence, minimal communication cost at each learning iteration, minimal low-overhead and privacy-preserving information exchange, i.e. only aggregated plans are exchanged between agents. This cost-effectiveness is feasible because of the tree communication structure~\cite{Petcu2005} that can be used to perform (i) efficient aggregation of the selected plans ($\numagents-1$ exchanged messages) and (ii) incremental decision-making used for coordination, i.e. an agent selects a plan based on the selected plans of its descendants.} as shown in comparisons with state of the art algorithms~\cite{pournaras2018,hinrichs2013,Hinrichs2017}. A high performance profile provides isolation of the performance analysis on structure, while inefficiencies by other algorithmic design choices are minimized. (ii) The core of the algorithm does not rely on exploration\footnote{Well-performing but suboptimum trapped solutions are found.}. Therefore the effect of structural self-adaptation as an exploration strategy can be easier isolated and studied.

\section{Related Work}\label{sec:related-work}

There is evidence that structure plays a key role in exploration as certain topological properties support more effective communication and information diffussion~\cite{Mason2012}. Agents' coordinated communication in reinforcement learning can be dynamically adapted to regulate learning performance and communication cost in distributed constraint optimization problems~\cite{Zhang2013}. Similarly, performance of reinforcement learning can be improved by self-organizing agents into a supervisory network on top of an agents' learning network. The latter is structured in dynamically formed groups via agents' negotiation~\cite{Zhang2010}. Structure may change as a result of network uncertainties such as network failures, latency and limited computational resources. Loss of learning performance as a result of such structural changes can be mitigated by localizing the learning process in part of a surviving network~\cite{Pournaras2019d}. 

All aforementioned approaches involve (self-)organizational changes to improve learning performance, i.e. topological changes are performed. In contrast, this paper studies the agents' relative repositioning over a fixed network and how it influences learning performance. Therefore, the scope of this paper has a more foundational character, while the findings illustrated are expected to provide new insights on the design of self-organization mechanisms for collective intelligence. 

The agents' positioning in a fixed learning structure can also be seen as an initialization problem of machine learning algorithms such as choosing the number of clusters and initial centroids in k-means that may result in slow convergence and empty clusters~\cite{celebi2011}. Bootstrapping solutions have been studied in this context~\cite{bradley1998}. Oscillating neural networks around optimal solutions may be a result of exploding gradients caused by high initializing weights~\cite{bengio1994,pascanu2013}, instead of more random ones around zero~\cite{friedman2001}. It has also been shown that linearly non-separable data require hidden layers between input and output layers~\cite{haykin1994}, while more than three hidden layers do not improve the learning performance of a feed-forward multi-layer perceptron~\cite{karsoliya2012}. Finally, adaptation of a neural network structure to the dimensionality of training data is critical to generalize and prevent over-fitting~\cite{walczak1999}. Applying these findings in the context of multi-agent systems and decentralized learning over networks with uncertainties is challenging and subject of ongoing research. 

\section{Structural Self-adaptation}\label{sec:structural-self-adaptation}

Structural self-adaption is the repositioning of a set of agents \agentset in a fixed tree topology and can be formalized as a \emph{bijection of isomorphic graphs}:


\begin{definition}[\textbf{Bijection}]\label{def:bijection}
\emph{Let two tree graphs \graphG and \graphH having each a set of vertices $\vertexset{\graphG}$, $\vertexset{\graphH}$ and edges $\edgeset{\graphG}$, $\edgeset{\graphH}$ such that $\cardinality{\vertexset{\graphG}} = \cardinality{\vertexset{\graphH}} = \cardinality{\agentset} = \numagents$ and $\cardinality{\edgeset{\graphG}} = \cardinality{\edgeset{\graphH}} = \numagents - 1$. 
A bijection \bijection between the vertex sets of graphs \graphG and \graphH is defined as $\bijection:\vertexset{\graphG} \rightarrow \vertexset{\graphH}$ such that two vertices $v$ and $w$ are adjacent in \graphG if and only if vertices $\bijectionof{v}$ and $\bijectionof{w}$ are adjacent in \graphH.
}
\end{definition}

\begin{definition}[\textbf{Isomorphism}]\label{def:isomorphism}
\emph{A graph \graphH obtained by a bijection \bijection on graph \graphG is an isomorphic graph to $ \graphG \simeq \graphH$.}
\end{definition}

These definitions determine a fixed tree topology for the graphs \graphG and \graphH, whose vertices host agents in different relative positions. There are $\cardinality{\bijectionset} = \numagents !$ possible bijections, where \bijectionset denotes all possible bijections on \graphG. Two bijection types are distinguished: (i) \emph{Fixed}: These concern the agents' positioning according to deterministic criteria. This paper reviews 124 criteria derived from metrics that measure features of the agents' plans. They are used to sort agents in an ascending or descending order and position agents in the tree in a bottom-up breadth-first manner. Fixed bijections can be compared to show how different metrics with which sorting is performed influence learning performance. (ii) \emph{Dynamic}: These concern the random or deterministic repositioning of the agents during learning runtime as the means to perform exploration of the solutions space so that learning performance improves by escaping from suboptimal trapped solutions. Four design elements model dynamic bijections and they are used to construct two online structural self-adaptation strategies.

In a decentralized environment in which each agent has a partial view of other agents, bijections of isomorphic tree graphs can be applied with two approaches: (i) \emph{migration} and (ii) \emph{self-organization}. Migration is the transfer of a piece of software and its state from one host to another. Migrations are earlier applied to multicasting trees and wireless sensor networks~\cite{Gupta2015}. They have the advantage that no topological changes are required in the communication network, i.e. TCP connections are preserved. However, software migrations pose security challenges and may consume significant bandwidth resources. On the other hand, self-organization preserves the software locality, while the parent-child connections are adapted via a communication protocol. AETOS, the \emph{Adaptive Epidemic Tree Overlay Service} is an example of such a self-organization mechanism~\cite{pournaras2010}. Agents realize a bijection by interacting with each other to discover their new parent and children without any centralized mediating trusted party. A new pairing of two agents can be determined by their proximity e.g. Euclidean distance, between the \emph{ranking score} of the two agents. The ranking score is calculated by deterministic metrics, i.e. Table~\ref{table:metrics}, or by a random score assignment, in case of an exploration. Decentralized building and maintenance of tree topologies has been extensively studied in earlier work with applications covering multimedia multicasting~\cite{Yeo2004} and distributed databases~\cite{Risson2006}. In contrast, this paper focuses on (i) the impact of agents' repositioning on the learning performance by self-organization as well as (ii) how repositioning can be triggered to improve learning performance. These challenges are not addressed in earlier work. They are the subject and contributions of this paper.

\section{Fixed Bijections}\label{sec:fixed-bijections}

Fixed bijections are a subset of \bijectionset and they are applied by executing the following steps: (1) Determine agents' information based on which the ranking score is calculated. (2) Determine a metric to represent the agents' information of Step 1. (3) Calculate the ranking score of each agent with the metric of Step 2. (4) Reposition and sort agents in a bottom-up breadth-first manner according to their ranking score. 

In Step 1, agents may determine their ranking score based on the following: (i) the \emph{plans}, (ii) the \emph{plan costs} or (iii) their \emph{preferences} on the global vs. local objective. This paper focuses\footnote{Plan costs are often derived from the plans, therefore, the focus on the plans is more generic. Determining different agents' preferences biases learning performance, i.e. trade-offs between global and local objectives~\cite{pournaras2018}. Therefore, plan costs and preferences are left for future work.} on the plans to limit the number of studied dimensions.

Step 2 determines plan representations based on which isomorphic tree graphs are generated. Such representations are meta-features measuring deterministic plan characteristics. Table~\ref{table:metrics} introduces 62 metrics some of which include the following\footnote{$corr_{p}$, $corr_{k}$ and $corr_{s}$ denote the Pearson, Kendal and Spearman correlation coefficient respectively. DST 1, DST 2 and DST 3, and respectively $dst_{1}$, $dst_{2}$, $dst_{3}$, indicate discrete sine transformation of type 1, 2 and 3. Similarly, DCT 1, DCT 2 and DCT 3, and respectively $dct_{1}$, $dct_{2}$, $dct_{3}$, indicate discrete cosine transformation of type 1, 2 and 3.}: minimum and maximum value of the possible plans, several correlation coefficients, metrics based on discrete sine and cosine transformations of the possible plans as well as metrics based on the Fourier transformation. 

\begin{table*}[!htb]
\caption{Reference list of structural self-adaptation metrics used as meta-features to improve learning performance. The mean operator \averageover{\cdot}{\cdot} iterates over the subscript elements, while $\sigma(\cdot)$ indicates the standard deviation and $\cartesianproduct$ the Cartesian product.}\label{table:metrics} 
\centering
\vfill
\resizebox{\textwidth}{!}{%
\begin{tabular}{lll}
		\toprule
		Display Name & Full Name & Formula \\
		\toprule
		avg-stdev & \makecell[l]{average standard \\ deviation}	& $\averageover{\plan \in \planset{\agent}}{ \sigma \left(\plan\right) }$ \\
		max-stdev & \makecell[l]{maximal standard\\deviation}	& $\maximal{\plan \in \planset{\agent}}{ \sigma \left(\plan\right) }$ \\
		min-stdev & \makecell[l]{minimal standard\\deviation}	& $\minimal{\plan \in \planset{\agent}}{ \sigma \left(\plan\right) }$ \\		
		max-value & maximum value & $\maximal{\plan \in \planset{\agent}, j=1, ..., \plandim} \planat{j} $ \\
		min-value & minimum value & $\minimal{\plan \in \planset{\agent}, j=1, ..., \plandim} \planat{j} $ \\
		
		\midrule
		
		avg-corr-pearson & \makecell[l]{average Pearson\\coefficient} & $\averageover{ \plan_{1}, \plan_{2} \in \planset{\agent} \times \planset{\agent}}{ \averageover{j=1, ..., \plandim} {corr_{p}\left(\plan_{1}, \plan_{2}\right)_{j}}
		}$ \\
		avg-corr-kendall & \makecell[l]{average Kendall\\coefficient} & $\averageover{ \plan_{1}, \plan_{2} \in \planset{\agent} \times \planset{\agent}}{ \averageover{j=1, ..., \plandim} {corr_{k}\left(\plan_{1}, \plan_{2}\right)_{j}}
		}$ \\
		avg-corr-spearman & \makecell[l]{average Spearman\\coefficient} & $\averageover{ \plan_{1}, \plan_{2} \in \planset{\agent} \times \planset{\agent}}{ \averageover{j=1, ..., \plandim} {corr_{s}\left(\plan_{1}, \plan_{2}\right)_{j}}
		}$ \\
		
		\midrule
		
		max-avg-corr-pearson & \makecell[l]{max average Pearson\\coefficient} & $\maximal{\plan_{1}, \plan_{2} \in \planset{\agent} \times \planset{\agent} } \averageover{j=1, ..., \plandim} {corr_{p}\left(\plan_{1}, \plan_{2}\right)_{j}}$ \\
		max-avg-corr-kendall & \makecell[l]{max average Kendall\\coefficient} & $\maximal{\plan_{1}, \plan_{2} \in \planset{\agent} \times \planset{\agent} } \averageover{j=1, ..., \plandim} {corr_{k}\left(\plan_{1}, \plan_{2}\right)_{j}}$ \\
		max-avg-corr-spearman & \makecell[l]{max average Spearman\\coefficient} & $\maximal{\plan_{1}, \plan_{2} \in \planset{\agent} \times \planset{\agent} } \averageover{j=1, ..., \plandim} {corr_{s}\left(\plan_{1}, \plan_{2}\right)_{j}}$ \\
		
		\midrule
		
		min-avg-corr-pearson & \makecell[l]{min average Pearson\\coefficient} & $\minimal{\plan_{1}, \plan_{2} \in \planset{\agent} \times \planset{\agent} } \averageover{j=1, ..., \plandim} {corr_{p}\left(\plan_{1}, \plan_{2}\right)_{j}}$ \\
		min-avg-corr-kendall & \makecell[l]{min average Kendall\\coefficient} & $\minimal{\plan_{1}, \plan_{2} \in \planset{\agent} \times \planset{\agent} } \averageover{j=1, ..., \plandim} {corr_{k}\left(\plan_{1}, \plan_{2}\right)_{j}}$ \\
		min-avg-corr-spearman & \makecell[l]{min average Spearman\\coefficient} & $\minimal{\plan_{1}, \plan_{2} \in \planset{\agent} \times \planset{\agent} } \averageover{j=1, ..., \plandim} {corr_{s}\left(\plan_{1}, \plan_{2}\right)_{j}}$ \\
		
		\midrule
		
		avg-max-corr-pearson & \makecell[l]{average max Pearson\\coefficient} & $\averageover{\plan_{1}, \plan_{2} \in \planset{\agent} \times \planset{\agent}}{ \maximal{j=1, ..., \plandim} \quad corr_{p}\left(\plan_{1}, \plan_{2}\right)_{j} }$ \\
		avg-max-corr-kendall & \makecell[l]{average max Kendall\\coefficient} & $\averageover{\plan_{1}, \plan_{2} \in \planset{\agent} \times \planset{\agent}}{ \maximal{j=1, ..., \plandim} \quad corr_{k}\left(\plan_{1}, \plan_{2}\right)_{j} }$ \\
		avg-max-corr-spearman & \makecell[l]{average max Spearman\\coefficient} & $\averageover{\plan_{1}, \plan_{2} \in \planset{\agent} \times \planset{\agent}}{ \maximal{j=1, ..., \plandim} \quad corr_{s}\left(\plan_{1}, \plan_{2}\right)_{j} }$ \\
		
		\midrule
		
		avg-min-corr-pearson & \makecell[l]{average min Pearson\\coefficient} & $\averageover{\plan_{1}, \plan_{2} \in \planset{\agent} \times \planset{\agent}}{ \minimal{j=1, ..., \plandim} \quad corr_{p}\left(\plan_{1}, \plan_{2}\right)_{j} }$ \\
		avg-min-corr-kendall & \makecell[l]{average min Kendall\\coefficient} & $\averageover{\plan_{1}, \plan_{2} \in \planset{\agent} \times \planset{\agent}}{ \minimal{j=1, ..., \plandim} \quad corr_{k}\left(\plan_{1}, \plan_{2}\right)_{j} }$ \\
		avg-min-corr-spearman & \makecell[l]{average min Spearman\\coefficient} & $\averageover{\plan_{1}, \plan_{2} \in \planset{\agent} \times \planset{\agent}}{ \minimal{j=1, ..., \plandim} \quad corr_{s}\left(\plan_{1}, \plan_{2}\right)_{j} }$ \\
		
		\midrule
		
		max-corr-pearson & \makecell[l]{max Pearson\\coefficient} & $\maximal{\plan_{1}, \plan_{2} \in \planset{\agent} \times \planset{\agent}}{ \minimal{j=1, ..., \plandim} \quad corr_{p}\left(\plan_{1}, \plan_{2}\right)_{j} }$ \\
		max-corr-kendall & \makecell[l]{max Kendall\\coefficient} & $\maximal{\plan_{1}, \plan_{2} \in \planset{\agent} \times \planset{\agent}}{ \minimal{j=1, ..., \plandim} \quad corr_{k}\left(\plan_{1}, \plan_{2}\right)_{j} }$ \\
		max-corr-spearman & \makecell[l]{max Spearman\\coefficient} & $\maximal{\plan_{1}, \plan_{2} \in \planset{\agent} \times \planset{\agent}}{ \minimal{j=1, ..., \plandim} \quad corr_{s}\left(\plan_{1}, \plan_{2}\right)_{j} }$ \\
		
		\midrule
		
		min-corr-pearson & \makecell[l]{min Pearson\\coefficient} & $\minimal{\plan_{1}, \plan_{2} \in \planset{\agent} \times \planset{\agent}}{ \minimal{j=1, ..., \plandim} \quad corr_{p}\left(\plan_{1}, \plan_{2}\right)_{j} }$ \\
		min-corr-kendall & \makecell[l]{min Kendall\\coefficient} & $\minimal{\plan_{1}, \plan_{2} \in \planset{\agent} \times \planset{\agent}}{ \minimal{j=1, ..., \plandim} \quad corr_{k}\left(\plan_{1}, \plan_{2}\right)_{j} }$ \\
		min-corr-spearman & \makecell[l]{min Spearman\\coefficient} & $\minimal{\plan_{1}, \plan_{2} \in \planset{\agent} \times \planset{\agent}}{ \minimal{j=1, ..., \plandim} \quad corr_{s}\left(\plan_{1}, \plan_{2}\right)_{j} }$ \\
		
		\midrule
		
		avg-dct1-coeff & \makecell[l]{average DCT 1\\coefficient} & $\averageover{\plan \in \planset{\agent}}{ \averageover{j=1, ..., \plandim}{ dct_{1}\left(\plan\right)_{j}}}$ \\
		avg-dct2-coeff & \makecell[l]{average DCT 2\\coefficient} & $\averageover{\plan \in \planset{\agent}}{ \averageover{j=1, ..., \plandim}{ dct_{2}\left(\plan\right)_{j}}}$ \\
		avg-dct3-coeff & \makecell[l]{average DCT 3\\coefficient} & $\averageover{\plan \in \planset{\agent}}{ \averageover{j=1, ..., \plandim}{ dct_{3}\left(\plan\right)_{j}}}$ \\
		
		\midrule
		
		max-dct1-coeff & max DCT 1 coefficient & $\maximal{\plan \in \planset{\agent}} \quad \maximal{j=1, ..., \plandim} \quad dct_{1}\left(\plan\right)_{j} $ \\


		\bottomrule								 
\end{tabular}
\begin{tabular}{lll}
		\toprule
		Display Name & Full Name & Formula \\
		\toprule
		
		max-dct2-coeff & max DCT 2 coefficient & $\maximal{\plan \in \planset{\agent}} \quad \maximal{j=1, ..., \plandim} \quad dct_{2}\left(\plan\right)_{j} $ \\
		max-dct3-coeff & max DCT 3 coefficient & $\maximal{\plan \in \planset{\agent}} \quad \maximal{j=1, ..., \plandim} \quad dct_{3}\left(\plan\right)_{j} $ \\
		
		\midrule			
		min-dct1-coeff & min DCT 1 coefficient & $\minimal{\plan \in \planset{\agent}} \quad \minimal{j=1, ..., \plandim} \quad dct_{1}\left(\plan\right)_{j} $ \\
		min-dct2-coeff & min DCT 2 coefficient & $\minimal{\plan \in \planset{\agent}} \quad \minimal{j=1, ..., \plandim} \quad dct_{2}\left(\plan\right)_{j} $ \\
		min-dct3-coeff & min DCT 3 coefficient & $\minimal{\plan \in \planset{\agent}} \quad \minimal{j=1, ..., \plandim} \quad dct_{3}\left(\plan\right)_{j} $ \\
		\midrule	
		
		avg-max-dct1-coeff & \makecell[l]{average max DCT 1\\coefficient} & $ \averageover{\plan \in \planset{\agent}}{ \maximal{j=1, ..., \plandim} \quad dct_{1}\left(\plan\right)_{j}}$ \\
		avg-max-dct2-coeff & \makecell[l]{average max DCT 2\\coefficient} & $ \averageover{\plan \in \planset{\agent}}{ \maximal{j=1, ..., \plandim} \quad dct_{2}\left(\plan\right)_{j}}$ \\
		avg-max-dct3-coeff & \makecell[l]{average max DCT 3\\coefficient} & $ \averageover{\plan \in \planset{\agent}}{ \maximal{j=1, ..., \plandim} \quad dct_{3}\left(\plan\right)_{j}}$ \\
		
		\midrule
		
		avg-min-dct1-coeff & \makecell[l]{average min DCT 1\\coefficient} & $ \averageover{\plan \in \planset{\agent}}{ \minimal{j=1, ..., \plandim} \quad dct_{1}\left(\plan\right)_{j}}$ \\
		avg-min-dct2-coeff & \makecell[l]{average min DCT 2\\coefficient} & $ \averageover{\plan \in \planset{\agent}}{ \minimal{j=1, ..., \plandim} \quad dct_{2}\left(\plan\right)_{j}}$ \\
		avg-min-dct3-coeff & \makecell[l]{average min DCT 3\\coefficient} & $ \averageover{\plan \in \planset{\agent}}{ \minimal{j=1, ..., \plandim} \quad dct_{3}\left(\plan\right)_{j}}$ \\
		
		\midrule
		
		avg-dst1-coeff & \makecell[l]{average DST 1\\coefficient} & $\averageover{\plan \in \planset{\agent}}{ \averageover{j=1, ..., \plandim}{ dst_{1}\left(\plan\right)_{j}}}$ \\
		avg-dst2-coeff & \makecell[l]{average DST 2\\coefficient} & $\averageover{\plan \in \planset{\agent}}{ \averageover{j=1, ..., \plandim}{ dst_{2}\left(\plan\right)_{j}}}$ \\
		avg-dst3-coeff & \makecell[l]{average DST 3\\coefficient} & $\averageover{\plan \in \planset{\agent}}{ \averageover{j=1, ..., \plandim}{ dst_{3}\left(\plan\right)_{j}}}$ \\
		
		\midrule
		
		max-dst1-coeff & max DST 1 coefficient & $\maximal{\plan \in \planset{\agent}} \quad \maximal{j=1, ..., \plandim} \quad dst_{1}\left(\plan\right)_{j} $ \\
		max-dst2-coeff & max DST 2 coefficient & $\maximal{\plan \in \planset{\agent}} \quad \maximal{j=1, ..., \plandim} \quad dst_{2}\left(\plan\right)_{j} $ \\
		max-dst3-coeff & max DST 3 coefficient & $\maximal{\plan \in \planset{\agent}} \quad \maximal{j=1, ..., \plandim} \quad dst_{3}\left(\plan\right)_{j} $ \\
		
		\midrule
		
		min-dst1-coeff & min DST 1 coefficient & $\minimal{\plan \in \planset{\agent}} \quad \minimal{j=1, ..., \plandim} \quad dst_{1}\left(\plan\right)_{j} $ \\
		min-dst2-coeff & min DST 2 coefficient & $\minimal{\plan \in \planset{\agent}} \quad \minimal{j=1, ..., \plandim} \quad dst_{2}\left(\plan\right)_{j} $ \\
		min-dst3-coeff & min DST 3 coefficient & $\minimal{\plan \in \planset{\agent}} \quad \minimal{j=1, ..., \plandim} \quad dst_{3}\left(\plan\right)_{j} $ \\
		
		\midrule
		
		avg-max-dst1-coeff & \makecell[l]{average max DST 1\\coefficient} & $ \averageover{\plan \in \planset{\agent}}{ \maximal{j=1, ..., \plandim} \quad dst_{1}\left(\plan\right)_{j}}$ \\
		avg-max-dst2-coeff & \makecell[l]{average max DST 2\\coefficient} & $ \averageover{\plan \in \planset{\agent}}{ \maximal{j=1, ..., \plandim} \quad dst_{2}\left(\plan\right)_{j}}$ \\
		avg-max-dst3-coeff & \makecell[l]{average max DST 3\\coefficient} & $ \averageover{\plan \in \planset{\agent}}{ \maximal{j=1, ..., \plandim} \quad dst_{3}\left(\plan\right)_{j}}$ \\
		
		\midrule
		
		avg-min-dst1-coeff & \makecell[l]{average min DST 1\\coefficient} & $ \averageover{\plan \in \planset{\agent}}{ \minimal{j=1, ..., \plandim} \quad dst_{1}\left(\plan\right)_{j}}$ \\
		avg-min-dst2-coeff & \makecell[l]{average min DST 2\\coefficient} & $ \averageover{\plan \in \planset{\agent}}{ \minimal{j=1, ..., \plandim} \quad dst_{2}\left(\plan\right)_{j}}$ \\
		avg-min-dst3-coeff & \makecell[l]{average min DST 3\\coefficient} & $ \averageover{\plan \in \planset{\agent}}{ \minimal{j=1, ..., \plandim} \quad dst_{3}\left(\plan\right)_{j}}$ \\
		
		\midrule
		
		sum-of-0-dft-coeff & \makecell[l]{sum of $0^{th}$ DFT\\coefficient} & $\abs{ \sum_{\plan \in \planset{\agent}} \fourierof{\plan}_{0} }$ \\
		max-of-0-dft-coeff & \makecell[l]{max of $0^{th}$ DFT\\coefficient} & $\abs{ \maximal{\plan \in \planset{\agent}} \quad \fourierof{\plan}_{0} }$ \\
		sum-non0-dft-coeff & \makecell[l]{sum of non-$0^{th}$ DFT\\coefficient} & $\abs{ \sum_{\plan \in \planset{\agent}} \sum_{j=1, ..., \plandim} \fourierof{\plan}_{j} }$ \\
		max-non0-dft-coeff & \makecell[l]{max of non-$0^{th}$ DFT\\coefficient} & $\abs{ \maximal{\plan \in \planset{\agent}} \quad \maximal{j=1, ..., \plandim} \quad \fourierof{\plan}_{j} }$ \\
		sum-all-dft-coeff & sum of DFT coefficient & $\abs{\sum_{\plan \in \planset{\agent}} \sum_{j=1, ..., \plandim} \fourierof{\plan}_{j}} $ \\
		avg-stdev-dft-coeff & \makecell[l]{average std dev\\of DFT coefficients} & $\abs{ \averageover{\plan \in \planset{\agent}}{ \sigma \left( \fourierof{\plan} \right) } }$ \\
		\bottomrule								 
\end{tabular}
}
\vspace{-0.35cm}
\end{table*}

Next in Step 3, agents use the selected metrics to calculate their ranking score with which a proximity between agent pairs can be determined, e.g. Euclidean distance.

\begin{figure}[!htb]
\vspace{-0.3cm}
\centering	
\includegraphics[width=0.45\columnwidth]{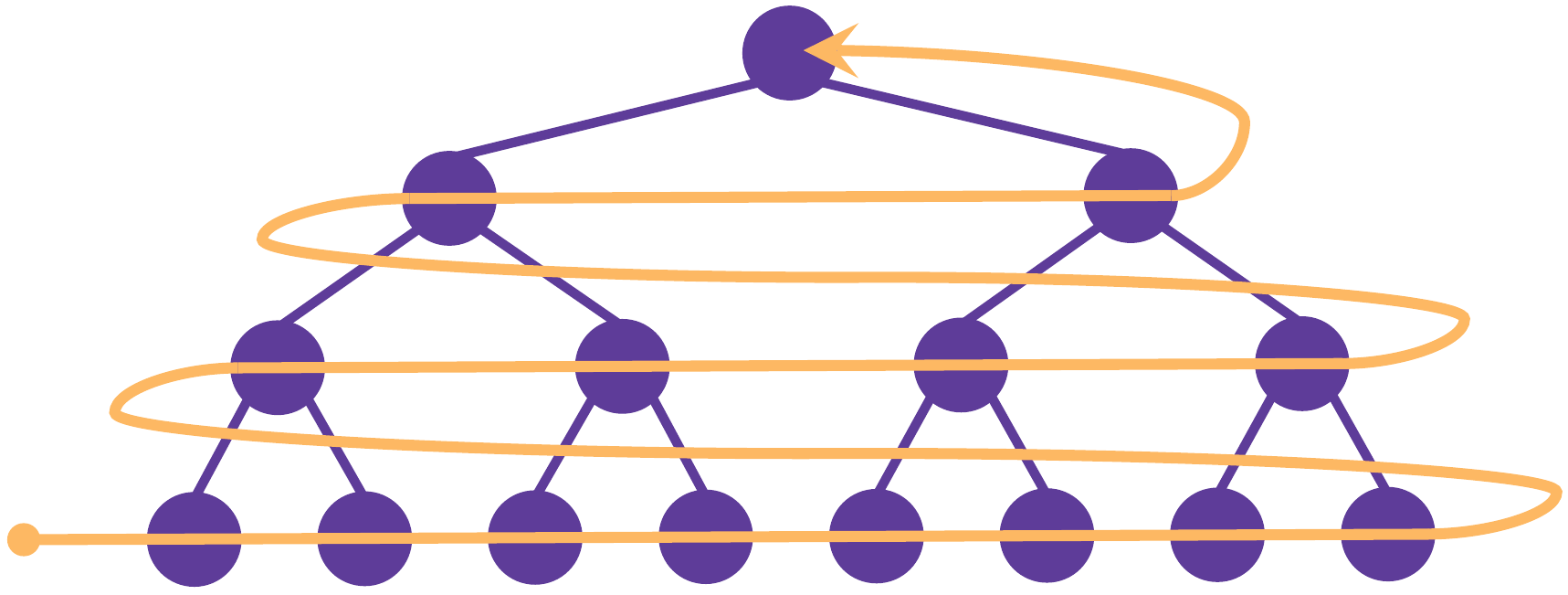}\caption{Bottom-up, breadth-first in a balanced tree.}\label{agent-assignments:tree-population}	
\vspace{-0.3cm}
\end{figure}

Finally, in Step 4, self-organization mechanisms, such as AETOS~\cite{pournaras2010}, use the agents' proximity information to reposition agents on the same balanced tree by sorting them in a bottom-up breadth-first manner as shown in Figure~\ref{agent-assignments:tree-population}. Agents can be positioned in both ascending and descending order.

To incept the impact of fixed bijections on learning performance consider a sequence of consecutive agents' decisions that progresses from the leaves up to the root. Each agent that performs its plan selection coordinates with the other agents underneath, i.e. takes into account the aggregate selected plans of these agents. Assume an agent with highly influential plans in the sense that when these plans aggregate with the plans of other agents, the global cost explodes. This can be because of plans with extreme oscillations or plans, whose mean value is significantly higher than the one of the other agents. The closer to the root this agent with the influential plans is, the lower the likelihood to adjust the aggregate selections of the agents underneath. This is because the number of the remaining agents above decreases and the remaining plan selections may not be enough to lower down the global cost. Therefore, learning performance can potentially improve if this agent is positioned at the bottom part of the tree to preserve a higher number of agents above to compensate and eventually drop down the global cost.

\section{Dynamic Bijections}\label{sec:dynamic-bijections}

A structural self-adaptation via a bijection initializes a new \emph{learning phase} in which learning performance improves further. Four design elements are introduced to model strategies for structural self-adaptations: (i) A mechanism to realize a bijection. (ii) A memory scheme of solutions for initializing the next learning phase. (iii) A criterion to trigger structural self-adaptation. (iv) A bijection type to apply. 

\subsection{Realizing a bijection}\label{subsec:bijection-realization}

Agents can realize themselves a structural self-adaptation without relying on a trusted third party. Assuming agents can execute a migration~\cite{Gupta2015} or self-organization service such as AETOS~\cite{pournaras2010}, this execution can be independently triggered by each I-EPOS agent based on system-wide criteria. For instance, the global cost, i.e. variance, is calculated using the final aggregate plan propagated during the top-down phase of each learning iteration. This approach assumes common criteria for all agents, otherwise agents need to reach consensus of when to perform the bijection. Such consensus can be reached via, for instance, voting mechanisms, which are broadly used in distributed ledgers and blockchain technologies~\cite{Mingxiao2017}. 

A migration or self-organization service can also been seen as a black box: Each agent provides as input its ranking score as well as its current children and parent, while it receives as output the respective new ones. Aggregate plans become outdated by establishing a new structure. Descendants change and the learning process requires reinitialization. Nevertheless, the selected plans at an earlier learning phase remain valid and represent a (good) found solution. Therefore, the learning process with the new agents' positioning can be initialized with these plans, instead of default random ones, to explore whether this previous solution can be further improved while preventing performance degradation. Structural self-adaptation can be triggered multiple times to explore the solution space until one or more conditions are met, e.g. the global cost is lower than some threshold or until a maximal number of learning iterations is performed. 

\subsection{Reinitialization via short-term vs. long-term memory}\label{sec:memory-schemes}

After structural self-adaptation, a new learning process is initialized with selected plans from the earlier learning phase. A (i) \emph{short-term} and (ii) \emph{long-term} memory scheme are introduced that assume limited resources: memory is bounded to the storage of a single earlier selected plan.

Short-term memory restores the selected plans from the \emph{last iteration} of the earlier learning phase. Long term memory restores the selected plans from an iteration of the earlier phase determined by a fixed \emph{memory offset}. This offset is the learning iteration on which the selected plans are memorized. For instance, agents with an offset of 3, memorize the selected plans at the $3^{rd}$ iteration, which are restored at iteration $0$ of the next learning phase. An offset equals to the convergence iteration is equivalent to short-term memory. Different offset values are studied in this paper provided as learning parameters. 

While short-term memory memorizes the selected plans of a suboptimal trapped solution found on convergence, long-term memory sacrifices\footnote{This performance sacrifice can be prevented by memorizing selected plans from a larger number of iterations.} learning performance as the means to escape from this trapped solution and potentially discover a better performing one at the next learning phase. Between consecutive learning phases with offset $>0$, the global cost is monotonically non-increasing: at least as low as the one of the offset iteration during the previous learning phase.


\subsection{Criteria to trigger structural self-adaptation}\label{sec:-trigger-criteria}

Two criteria for triggering structural self-adaptation are introduced: (i) \emph{convergence} and (ii) \emph{global cost reduction}. 


The convergence criterion triggers structural self-adaptation when convergence is reached, i.e. when the global cost in two consecutive iterations remains the same. The criterion of global cost reduction triggers a structural self-adaptation when global cost drops below a certain \emph{threshold}, provided as system parameter. This threshold represents a sufficient decrease of the global cost measured by the \emph{slope}. The higher the slope, the higher the global cost reduction. Consequently, larger reduction steps are still in progress and further learning iterations are required for convergence. On the contrary, when the slope is low, global cost reduction decreases and as a result learning approaches convergence. The slope is measured by the relative difference between the global cost \globalcostat{\ttime} and \globalcostat{\ttime - 1} of two consecutive learning iterations \ttime and $\ttime-1$: $s = \frac{\globalcostat{\ttime - 1} - \globalcostat{\ttime}}{\globalcostat{\ttime - 1}}$. The absolute difference between the global costs at two consecutive iterations $w = \globalcostat{\ttime - 1} - \globalcostat{\ttime}$ is referred to as a \textit{residual}. The slope and the respective threshold receive values in the range $[0, 1]$. A threshold value of 0 prohibits structural self-adaptation, while a value of 1 forces such self-adaptation every two iterations, given that a current and a previous iterations are required to compute a residual. 

\subsection{Bijection types}\label{sec:bijection-types}

Two bijection types are distinguished: (i) \emph{deterministic} and (ii) \emph{random}. Deterministic bijections are the ones generated with metrics such as the ones of Table~\ref{table:metrics}. These metrics calculate the ranking score of the agents based on which sorting is performed. Determining the effectiveness of deterministic bijections in online structural self-adaptations is not straightforward as it highly depends on the memory scheme employed as well as the shape of the combinatorial landscape, i.e. the plans and their cost. Therefore, offline deterministic bijections are studied to understand their primitive role on learning performance before moving to an online context that is subject of future work. Random bijections are generated by assigning a random ranking score to each agent. Based on this random score, the self-organization service returns a repositioning of the agents, a bijection $\bijection \in \bijectionset$. Random bijections are used during the learning runtime as an exploratory strategy to escape from suboptimum trapped solutions.

\section{Online Structural Self-adaptation}\label{sec:strategies}

\begin{figure}[!htb]
\vspace{-0.35cm}
\centering
\subfloat[Convergence criterion with long-term memory.]{\includegraphics[width=0.49\columnwidth]{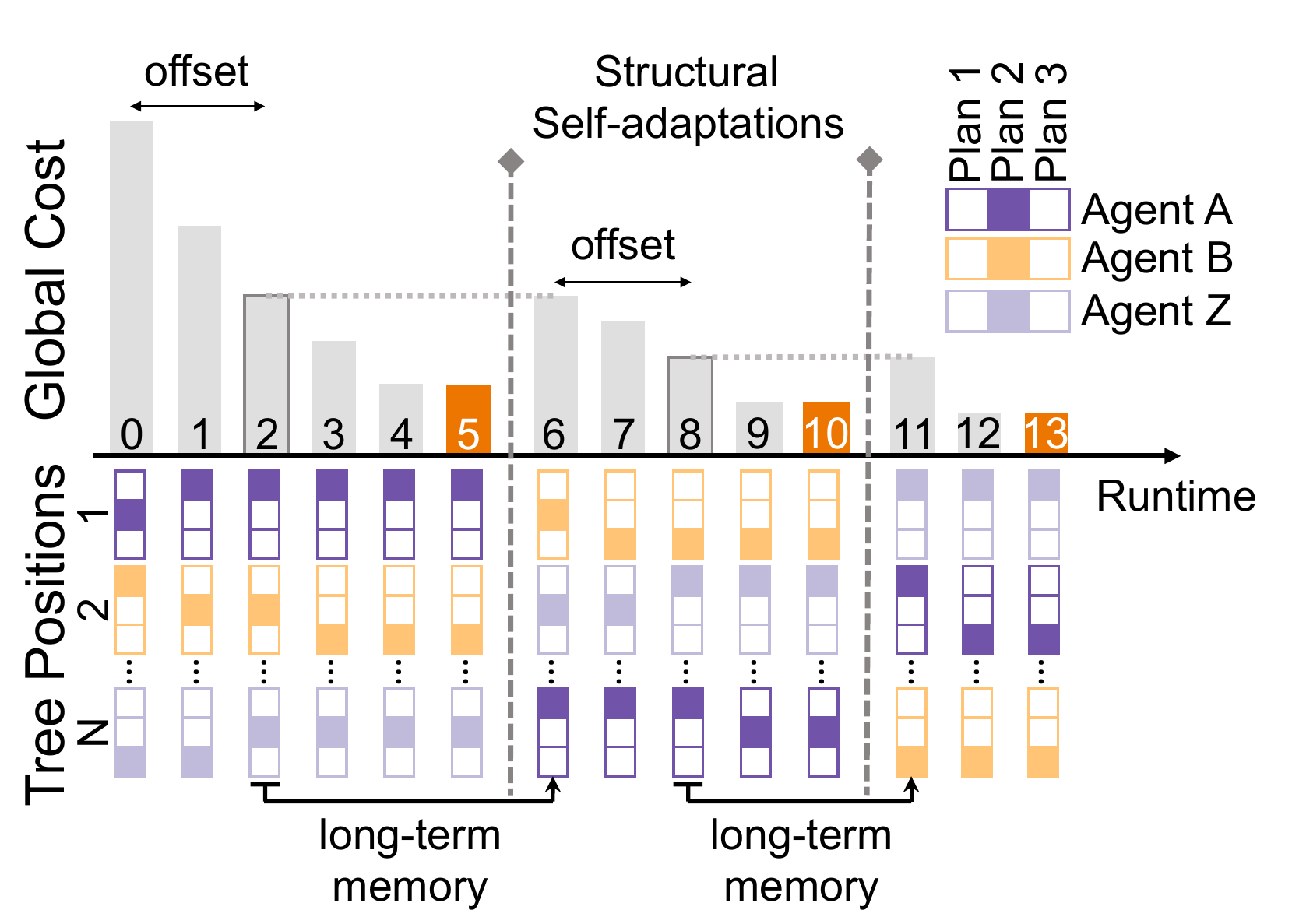}}\hfill
\subfloat[Global cost criterion with short-term memory.]{\includegraphics[width=0.49\columnwidth]{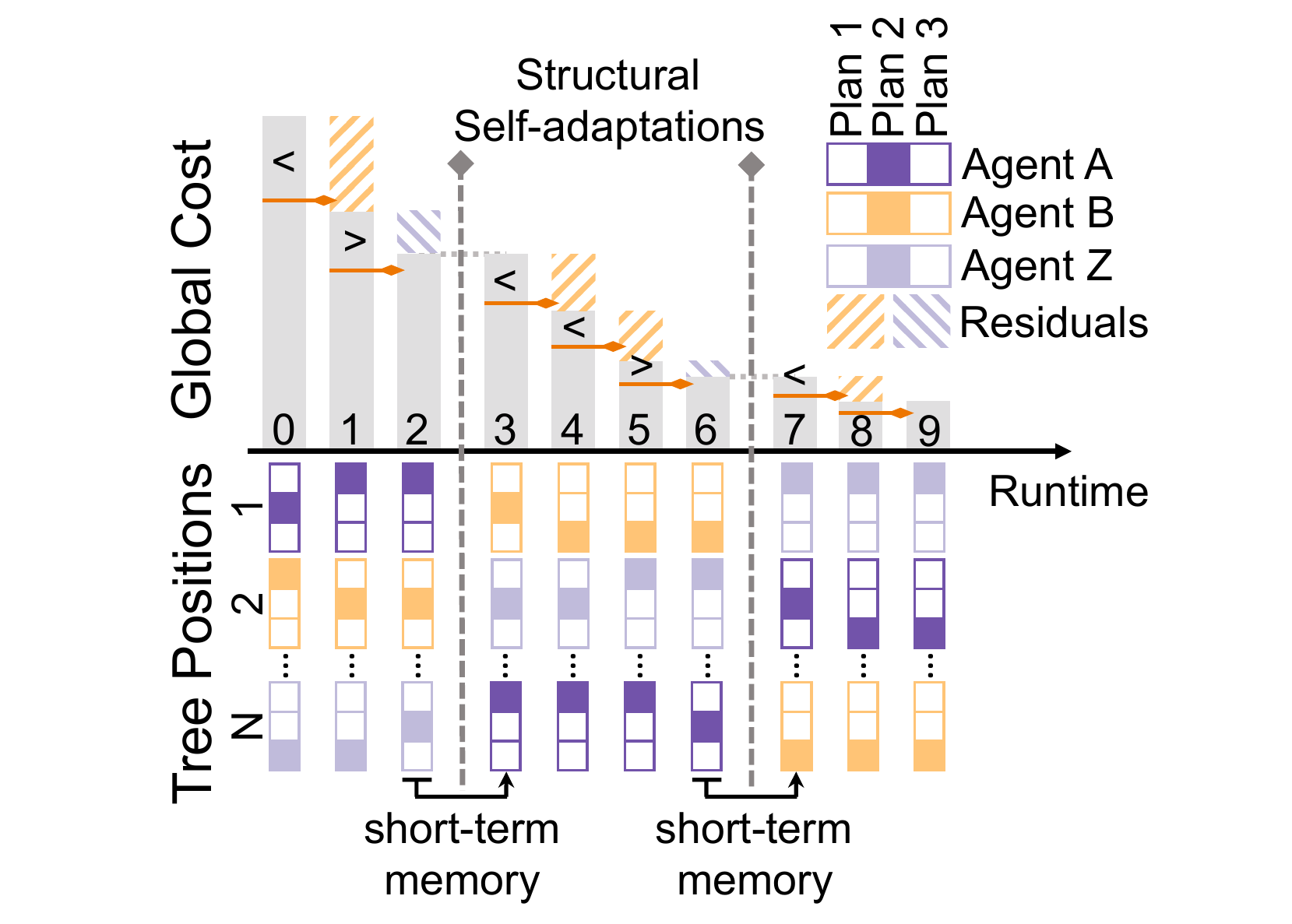}}	
\caption{The two online structural self-adaptation strategies.}\label{agent-strategies}
\vspace{-0.5cm}
\end{figure}

Figure~\ref{agent-strategies} introduces two online structural self-adaptation strategies based on dynamic bijections.

Figure~\ref{agent-strategies}a illustrates the strategy based on the convergence criterion with long-term memory. It is shown for an offset of $2$. Convergence is detected on the 5th iteration triggering structural self-adaptation. The new learning phase begins on the $6^{th}$ iteration initialized by the selected plans of the $2^{nd}$ iteration as indicated by the vertical dotted line. The process repeats until the $13{th}$ iteration when the algorithm terminates. The termination condition is the detection of convergence before reaching the offset iteration. 

Figure~\ref{agent-strategies}b shows the strategy based on the global cost reduction criterion with short-term memory. A threshold of 25\% is marked by horizontal pointers between the bars of two consecutive iterations. The residual on the 1st iteration is higher than the 25\% threshold that prevents self-adaptation. The slope drops below threshold on the $2^{nd}$ iteration triggering self-adaptation. The new learning phase is initialized on the $3^{rd}$ iteration with the selected plans of the $2^{nd}$ iteration indicated by the vertical dotted line. Note the preservation of the global cost reduction. The strategy criterion is met next on the $6^{th}$ iteration. Termination is detected on the $9^{th}$ iteration when no change in the slope is observed.

\section{Experimantal Methodology}\label{sec:methodology}

This section introduces the evaluation methodology of fixed and dynamic bijections. The employed synthetic and real-world datasets are discussed, followed by the parameterization of I-EPOS and the studied variables, i.e. number of children. A novel evaluation methodology is introduced for assessing the optimality of the solutions independent of the employed dataset. This methodology allows the systematic evaluation of the online strategies for structural self-adaptation.

\subsection{Synthetic and real-world datasets}\label{subsec:datasets}

Table~\ref{agent-assignment:benchmark-experimental-settings} outlines the datasets~\cite{Pournaras2019b} and the main experimental settings. A synthetic dataset with plan values drawn from a Normal distribution is used as well as two real-world datasets\footnote{A third real-world dataset is made available~\cite{Pournaras2019b}. It concerns the charging power consumption of electric vehicles and the planning methodology is introduced in earlier work~\cite{Pournaras2017b}. This paper focuses on the synthetic, energy and bicycle datasets due to space limitations.} from pilot projects~\cite{pournaras2018}: (i) \emph{Energy}--This dataset is a disaggregation result of the zonal power transmission system in the Pacific Northwest Smart Grid demonstration project~\cite{pournaras2017}. The first plan is the original disaggregated load. The following three plans are computed by the SHUFFLE plan generation scheme that randomly shuffles the values of the first plan. The next three plans are computed by the SWAP-15 generation scheme that randomly selects 15 pairs of values to swap. The final three plans are generated respectively via SWAP-30. Consequently, the mean and the standard deviation of all possible plans are equal for each agent. (ii) \emph{Bicycle}--This dataset consists of the trip records of the Hubway bike sharing system in Paris~\cite{pournaras2018}. Trips match to users via the data fields of zip code, year of birth and gender. Each plan value measures the difference in the number of incoming and outgoing trips at a bike station. For example, if a user cycled between stations 1 and 3, the corresponding possible plan is $\{-1, 0, 1, 0, ...\}$. Different trips of a user are encoded as different possible plans. 

\begin{table}[!htb]
\caption{An outline of the datasets and experimental settings.}\label{agent-assignment:benchmark-experimental-settings}
\centering
\resizebox{\columnwidth}{!}{%
\begin{tabular}{l l l l l}
\toprule
\multirow{1}{*}{Parameter} && \multicolumn{3}{c}{\textit{Datasets}~\cite{Pournaras2019b}} \\\cmidrule{3-5}
&&	Bicycle	& Energy & Synthetic \\\midrule
Global cost function &&	\multicolumn{3}{c}{minimization of variance} \\
Tree type	&&	\multicolumn{3}{c}{\emph{default}: binary $\graphH_{\bijection}, \quad \forall \bijection \in \setonemillion$}	\\\midrule		
Number of agents	&&	$\numagents = 1000$ &	$\numagents = 1000$	&	$\numagents = 1000$	\\
Number of plans		&&	$\cardinality{\planset{\agent}} \leq 23, \forall \agent \in \agentset$ &	$\cardinality{\planset{\agent}} = 10, \forall \agent \in \agentset$ &	$\cardinality{\planset{\agent}} = 16, \forall \agent \in \agentset$	\\
Dimension of plans	&&	$\plandim = 98$ &	$\plandim = 144$ &	$\plandim = 100$	\\
Number of iterations	&&	$T = 40$	&	$T = 40$ &	$T = 40$ \\\bottomrule
\end{tabular}
}
\vspace{-0.3cm}
\end{table}

%




%
%




\subsection{Collective learning parameterization}\label{subsec:learning-parameters}

I-EPOS runs by default on a binary balanced tree. A varied number of children, from 2 to 14, is evaluated for fixed bijections. A higher number of children for each agent results in more informed plan selections, however, the number of leaves that make plan selections without information from descendants increases: for a balanced binary tree with 1000 agents, the number of leaves is 500 (50\%), whereas for a 14-ary balanced tree, the number of leaves is 928 (92.8\%). I-EPOS minimizes the global cost function that is the variance used as a balancing criterion: reducing the power peaks and load-balancing the utilization of the bike sharing stations. To stretch the collective learning performance, only the global cost function is optimized and therefore the I-EPOS parameter of $\lambda=0$ ignores the local cost function. 


\subsection{Evaluation approach for optimality}\label{subsec:optimality-approach}

The empirical evaluation of optimality concerns the ranking of the found solution in terms of global cost out of all possible solutions in the combinatorial space that is $k^{\numagents}$. As the scale of the combinatorial space explodes for a high number of plans and agents, such an evaluation is particularly challenging. Earlier work limits the optimality evaluation to a low number of agents and plans per agent, e.g. $k^{\numagents}=2^{20}=4^{10}$~\cite{pournaras2018}. This paper contributes an alternative approach that constructs a representative sample of potentially\footnote{The assumption of sampling high-performing solutions is supported by evidence of the I-EPOS optimality when compared to brute-force search~\cite{pournaras2018}.} high-performing solutions using I-EPOS operating with random bijections applied on isomorphic tree graphs. This approach has the following advantages\footnote{The more naive approach of sampling solutions by random plan selections is not considered as these solutions are not result of any optimization in contrast to solutions derived by random bijections applied to I-EPOS.}: (i) A profiling of high-performing solutions for different datasets using the same learning methodology, I-EPOS, without employing other heuristics or brute-force. (ii) An efficient computation of a large number of random bijections and their learning performance is feasible and can be performed offline using parallel batch processing. 

A benchmark dataset is generated using the aforementioned approach with both deterministic and random bijections applied to I-EPOS. This dataset~\cite{Pournaras2019} is a contribution of this paper and it can be used to encourage and support further research on combinatorial optimization and learning. A total of 1 million ($\cardinality{\setonemillion} = 10^{6}$) random bijections\footnote{In total $1000!$ possible bijections that is roughly $4.024 \cdot 10^{2567}$.} are generated for each dataset. I-EPOS runs for 40 iterations with each of these bijections. In total, $10^{6}$ bijections $\ast$ $4$ datasets $\ast$ $40$ iterations $\ast$ $(1000-1)$ $\ast$ $2$ messages per iteration calculate the total of $\approx320$ \emph{billion learning messages}. I-EPOS instances run in parallel in their own Java virtual machine on a Hetzner dedicated server\footnote{20 parallel JVMs are deployed on a 3.5GHz CPU, 32GB RAM, 4TB HD Hetzner machine: https://www.hetzner.com 
(last accessed: March 2019). Each JVM runs I-EPOS with a random bijection. Execution lasted several months.}.

The performance profiling of the solutions discovered via random bijections depends on the application, i.e. the data values of the plans. The ranking of the solutions, according to global cost, can be generalized by calculating instead the \emph{percentile}\footnote{It represents how close the global cost, obtained with a certain isomorphic tree graph, is to the minimum one over the 1 million isomorphic tree graphs generated. Percentiles receive values in the range $[0,1]$. A percentile value can be placed in a 3-D vector space with each axis corresponding to a dataset. Isomorphic tree graphs closer to the origin of this space result in a lower overall global cost over the three datasets. The distance to the origin can be calculated with the Euclidean distance between the two points $(x_1, y_1, z_1)$ and $(x_2, y_2, z_2)$ as $\sqrt{(x_1 - x_2)^2 + (y_1 - y_2)^2 + (z_1 - z_2)^2}$, where the coordinates represent the global cost for each dataset.} in which a solution is found. This method assumes a common distribution of the global cost among the different datasets. To test this assumption, the global cost is modeled as a random variable \globalcostrandvar that has unknown distribution but is dependent on the random bijections \bijectionset. 

At first the probability density function of \globalcostrandvar is estimated in a \emph{non-parametric way}\footnote{Non-parametric estimation makes no assumptions on the underlying distribution, hence it has no parameters.} using the kernel density estimation fed with the global cost of I-EPOS for all isomorphic graphs of \setonemillion. A Gaussian kernel is used with a bandwidth suggested in earlier work~\cite{silverman1986}: $n^{-\frac{1}{5}}$, which is equal to $0.0631$ for $n = 10^{6}$. 

Next the global cost as a random variable \globalcostrandvar is modeled by a Gaussian distribution with unknown expectation and variance that are estimated using the mean ($50^{th}$ percentile) and variance of the global costs from the benchmark dataset. 

If the non-parametric kernel density estimation matches the parametric one, the learning performance based on percentiles can be reliably compared among the different datasets. With this methodology, the optimality of deterministic bijections can be rigorously studied by generalizing and linking their performance profile to the performance percentiles of random bijections. For each of the 62 metrics of Table~\ref{table:metrics}, two bijections are applied, each of them sorts agents in ascending and descending order. From the total of 124 fixed bijections, 39 of them are selected for illustration in this paper.


\subsection{Evaluation approach for self-adaptation strategies}\label{subsec:evaluation-strategies}

For the evaluation of the two structural self-adaptation strategies, I-EPOS is initialized with three random isomorphic tree networks corresponding to the $10^{th}$, $50^{th}$ and $90^{th}$ percentile referred to as \emph{baselines}, while a structural self-adaptation is triggered with a sample of an isomorphic tree graph from the whole set \bijectionset. Due to the random sampling involved, the execution of the two strategies repeats 100 times. Their learning performance is evaluated with the \textit{average relative improvement} in the global cost reduction between the baselines and the strategies over all repetitions. A positive relative improvement results in higher learning performance by the strategies compared to the baselines, while a negative one results in a lower learning performance respectively. The memory offset and the threshold for each of the two strategies vary in the range $[1,20]$ and $[0.1,0.9]$ respectively, with a step $0.1$ for the latter. The total learning runtime is 100 iterations during which structural self-adaptations are performed. 

Controlling the number of structural self-adaptations is critical for the cost-effectiveness of learning. Repositioning agents requires communication and computational cost, e.g. AETOS introduces interactions between agents to discover and connect the parent and child with the closest proximity~\cite{pournaras2010}. The two strategies are also compared in terms of the number of self-adaptation they perform until termination. 

\section{Experimantal Evaluation}\label{sec:evaluation}

This section validates the methodology for assessing learning optimality with isomorphic tree graphs followed by the evaluation of fixed and dynamic bijections. 

\subsection{Learning optimality with isomorphic tree graphs}\label{subsec:optimality}

\begin{figure}[!htb]
\centering	
\subfloat[Synthetic dataset]{\includegraphics[width=0.33\columnwidth]{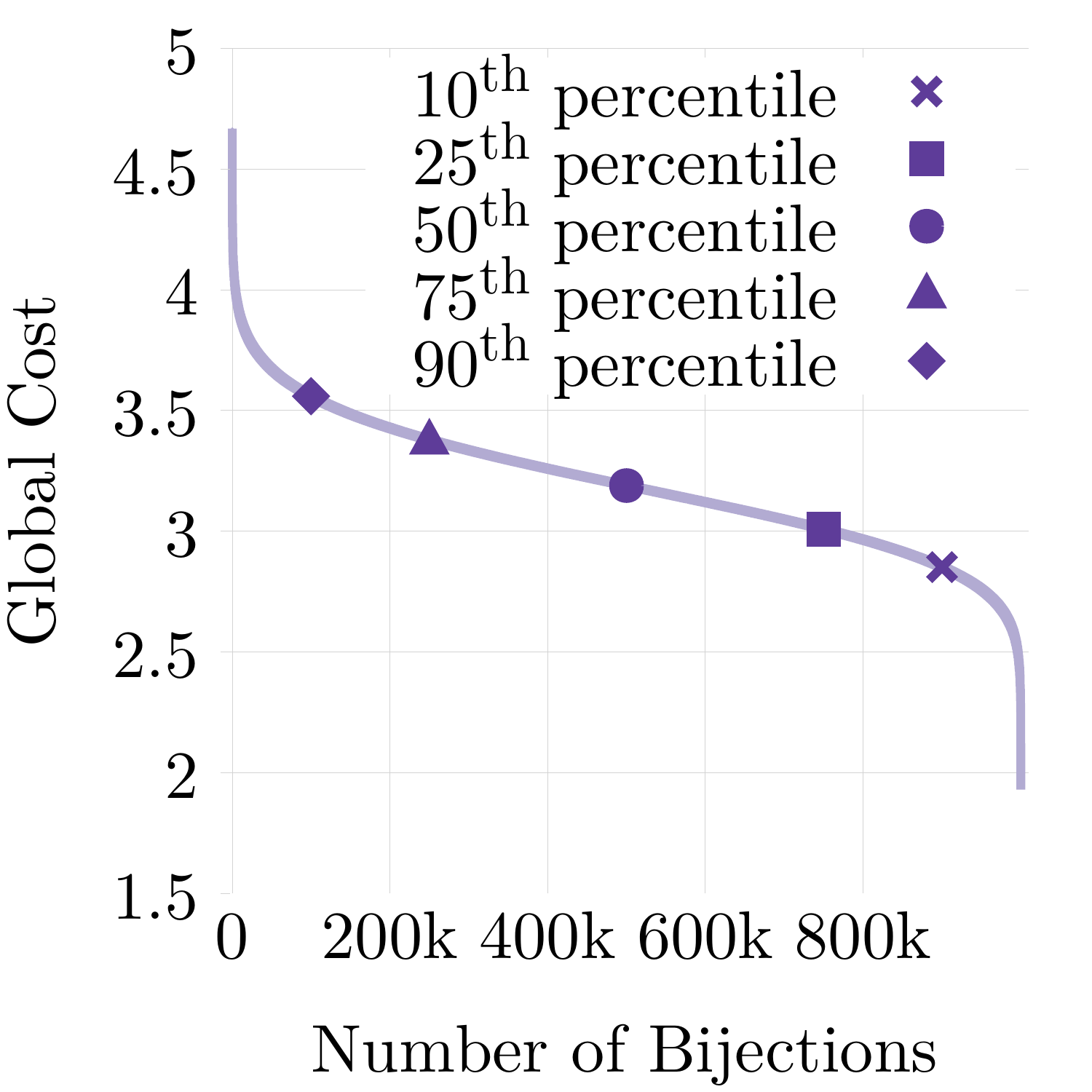}\label{benchmark:global-cost-curve:gaussian}}\hfill 
\subfloat[Energy dataset]{\includegraphics[width=0.33\columnwidth]{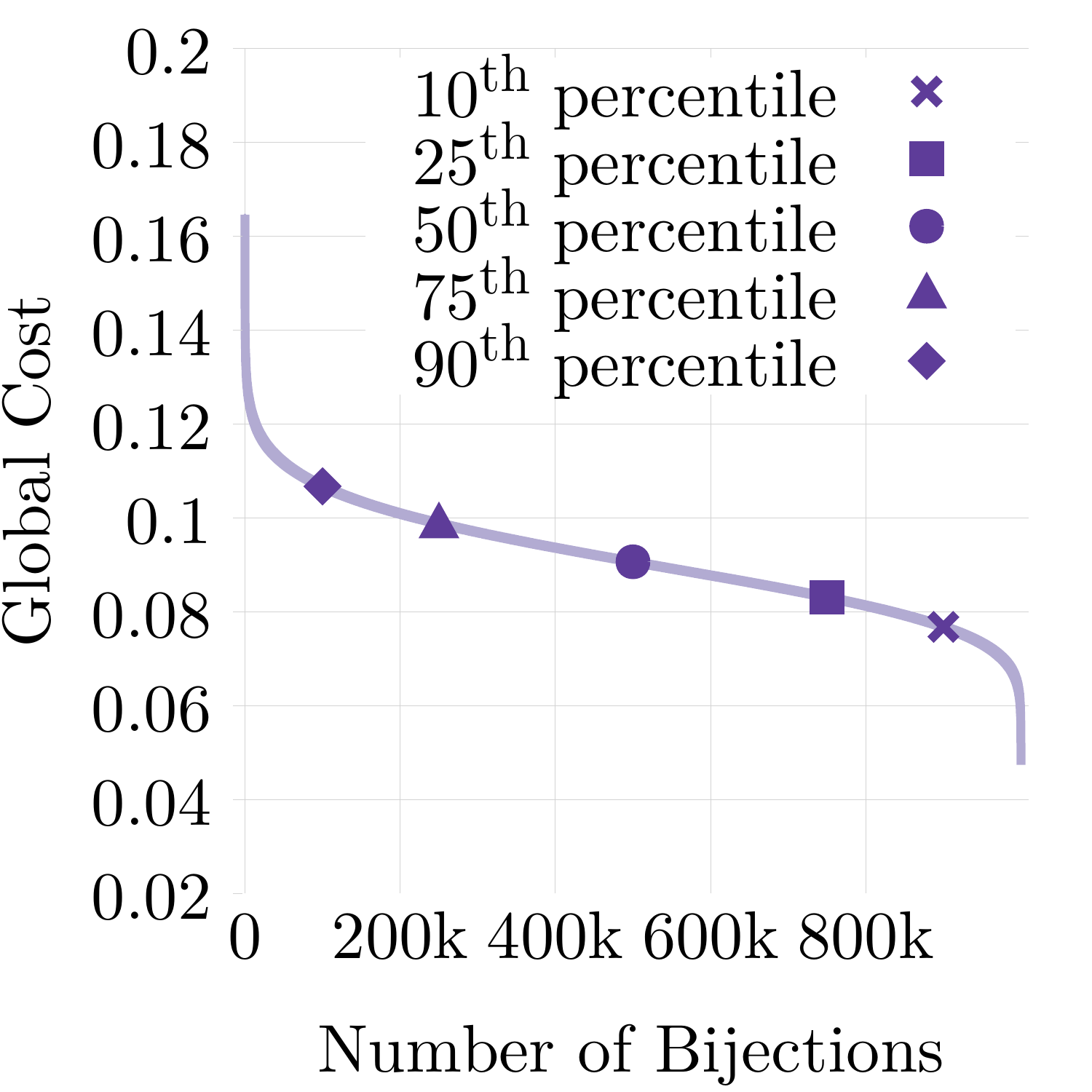}\label{benchmark:global-cost-curve:energy}}\hfill 
\subfloat[Bicycle dataset]{\includegraphics[width=0.33\columnwidth]{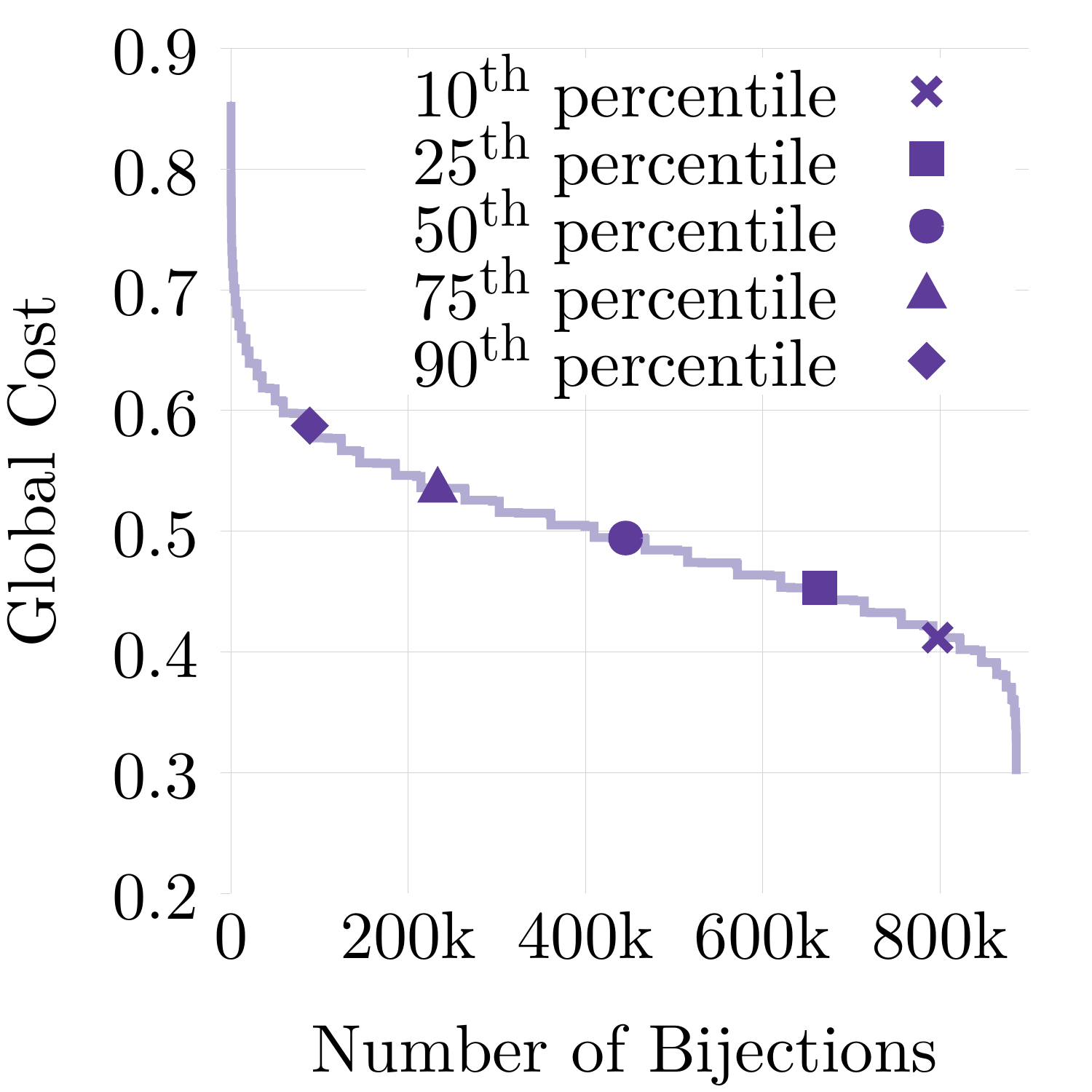}\label{benchmark:global-cost-curve:bicycle}}\hfill\\
\subfloat[synthetic dataset]{\includegraphics[width=0.33\linewidth]{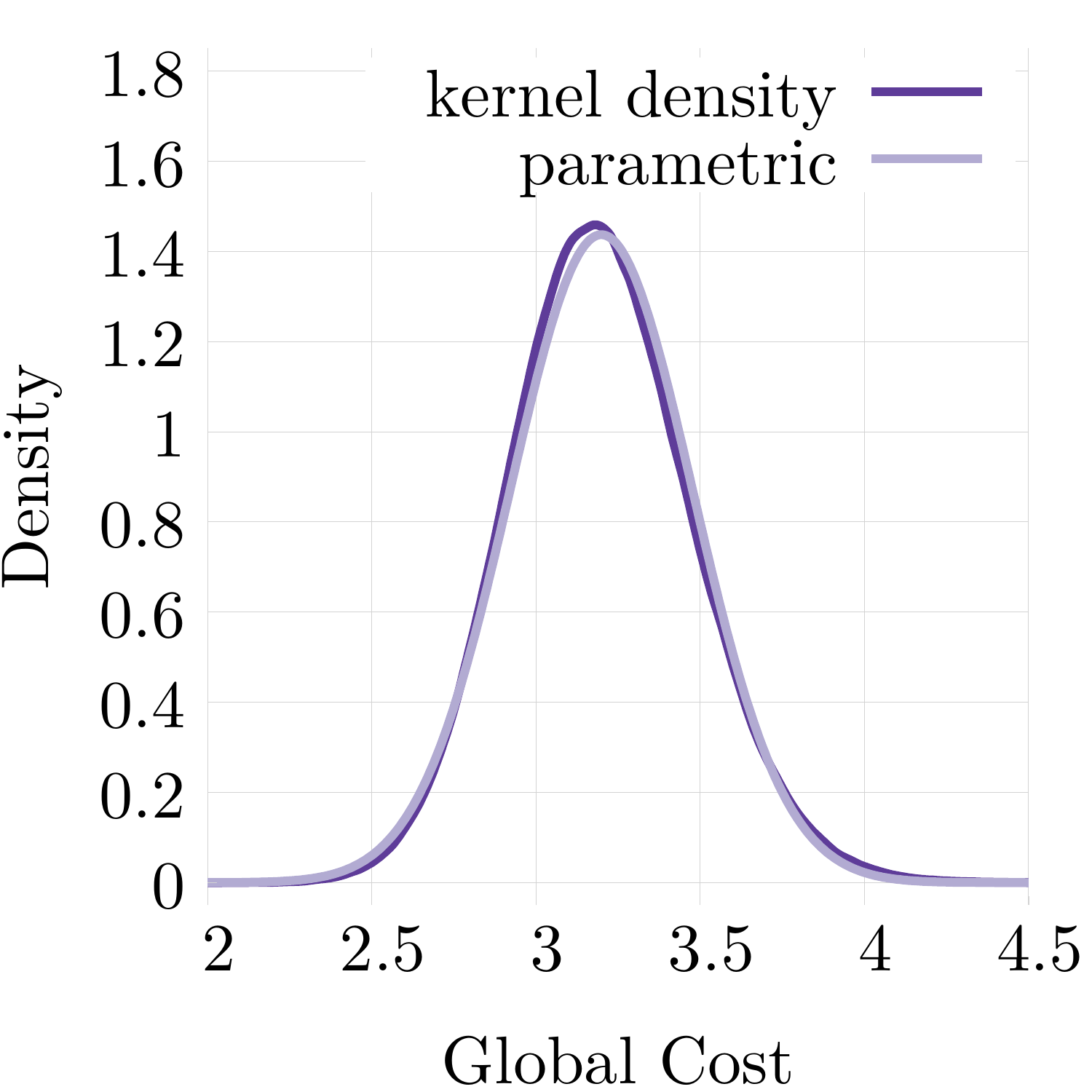}\label{benchmark:global-cost-density:gaussian}}\hfill 
\subfloat[Energy dataset]{\includegraphics[width=0.33\linewidth]{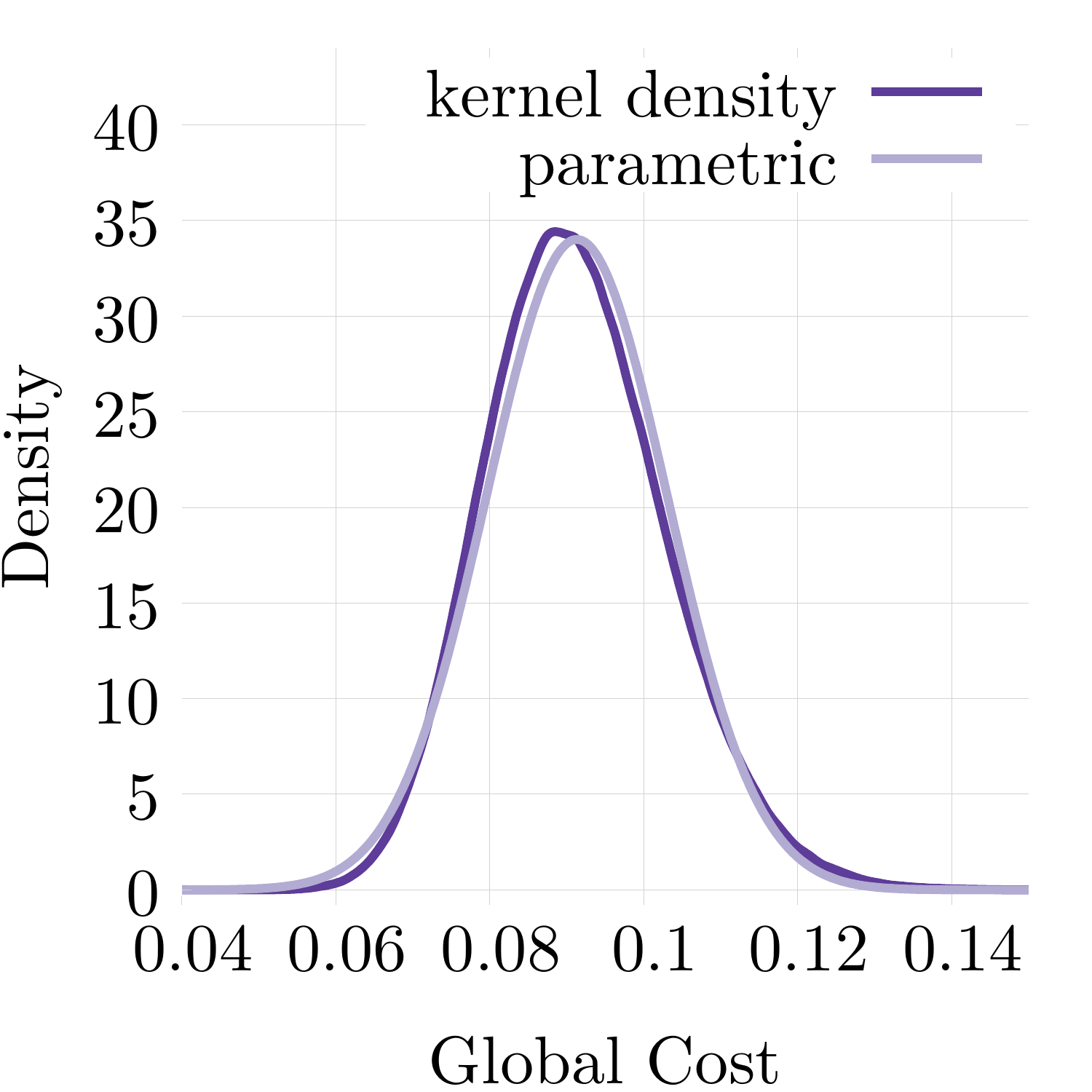}\label{benchmark:global-cost-density:energy}}\hfill 
\subfloat[Bicycle dataset]{\includegraphics[width=0.33\linewidth]{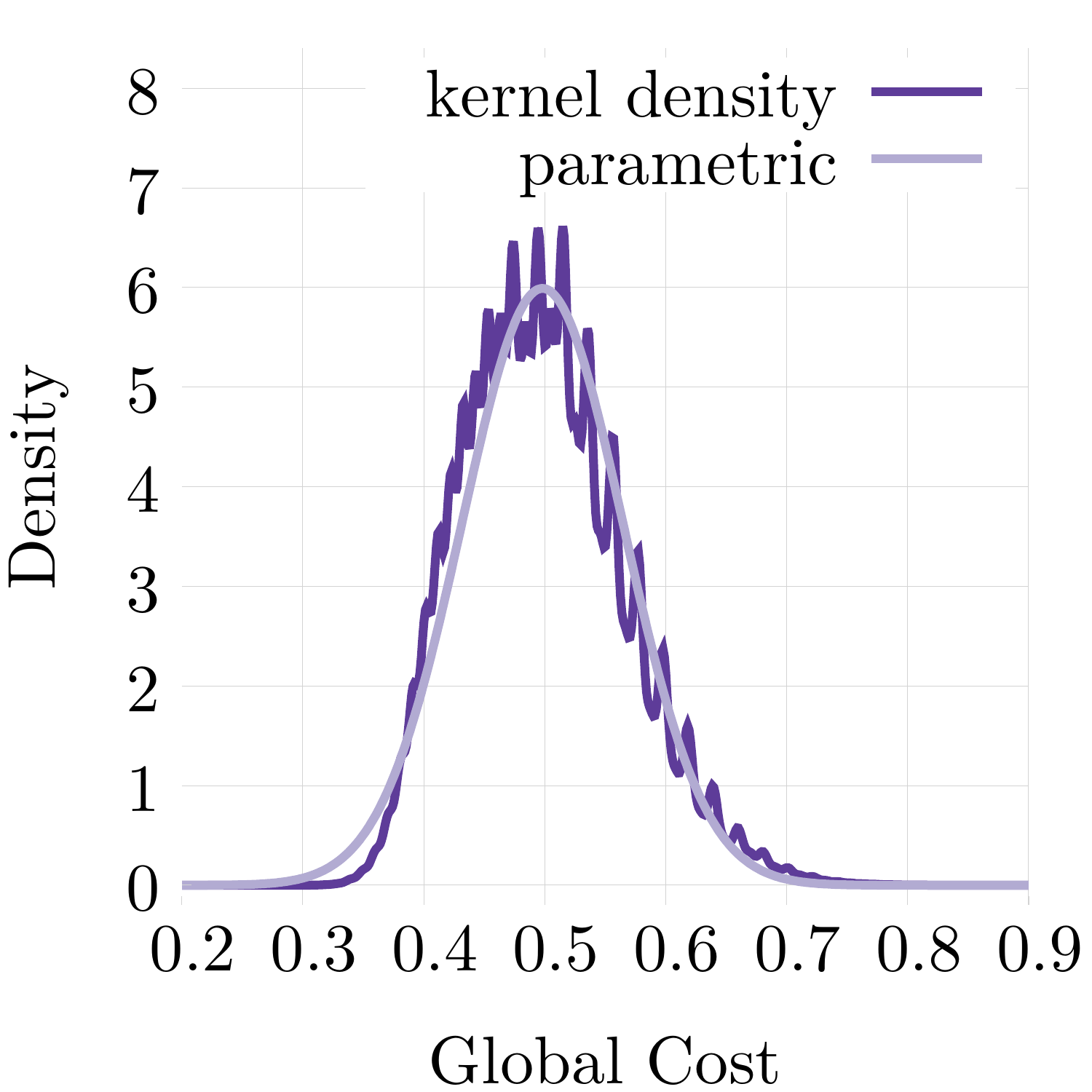}\label{benchmark:global-cost-density:bicycle}}\hfill
\caption{Performance profiling of $10^{6}$ random bijections.}\label{benchmark:global-cost-curves-densities}
\vspace{-0.3cm}
\end{figure}

The one million isomorphic graphs are sorted from high to low according to the global cost at convergence\footnote{On average, convergence occurs at the $13^{th}$, $24^{th}$, and $31^{st}$ iteration for the synthetic, energy and bicycle datasets respectively.} for each dataset and are shown in Figure~\ref{benchmark:global-cost-curve:gaussian},~\ref{benchmark:global-cost-curve:energy} and~\ref{benchmark:global-cost-curve:bicycle}. The following observations can be made: (i) The shape of all sorted solutions is similar among datasets, i.e. very few solutions at extremes while the majority decreases linearly. (ii) The solutions shape resembles the ones obtained for small-scale networks via brute-force as shown in Figure~21 and~22 of earlier work~\cite{pournaras2018}. (iii) There is accurate correspondence of the percentiles among the different datasets. (iv) The global cost reduction compared to the maximal observed value is 58.6\%, 71.2\%, 65.07\% for the synthetic, energy and bicycle dataset respectively.

The parametric and non-parametric global cost distributions obtained with random isomorphic graphs are compared in Figure~\ref{benchmark:global-cost-density:gaussian},~\ref{benchmark:global-cost-density:energy} and~\ref{benchmark:global-cost-density:bicycle} for each dataset. 

The expectation and variance of the parametric Gaussian distribution are estimated using the benchmark dataset constructed as follows: $\globalcostrandvar_{\text{s}} \sim \normaldist{\mu=3.1989}{\sigma^{2}=0.0771}$, $\globalcostrandvar_{\text{e}} \sim \normaldist{\mu=0.0913}{\sigma^{2}=0.0001}$ and $\globalcostrandvar_{\text{b}} \sim \normaldist{\mu=0.4981}{\sigma^{2}=0.0044}$ for the synthetic, energy and bicycle dataset respectively. The kernel density estimator approximates very closely the parametric densities in all three datasets without making any assumptions about the underlying data and therefore it holds that $\globalcostrandvar \sim \normaldist{\mu}{\sigma^{2}}$.

\subsection{Learning performance with fixed bijections}\label{subsec:fixed-bijections}

\begin{figure}[!htb]
\centering	
\includegraphics[width=\linewidth]{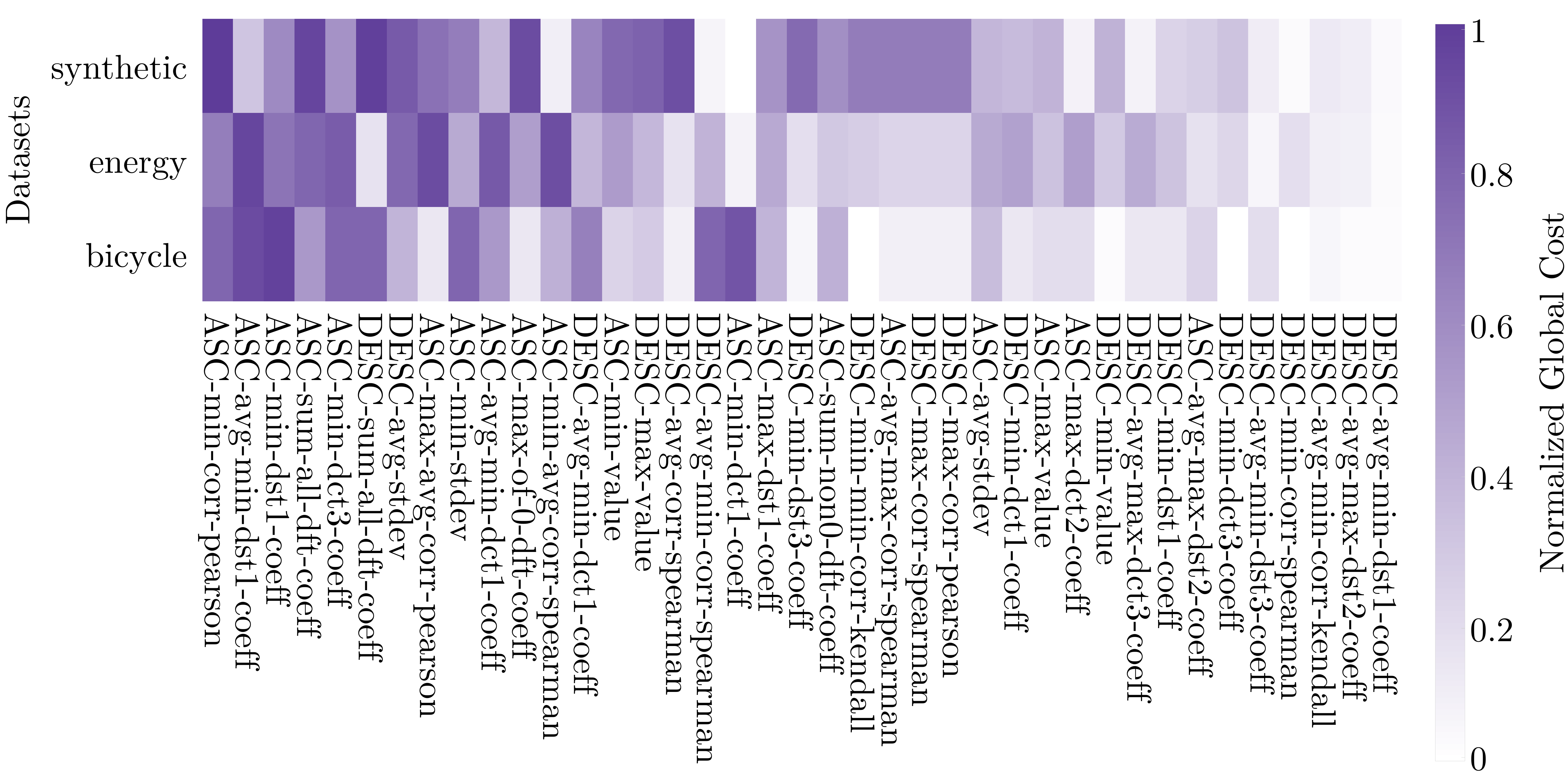}
\caption{Learning performance of selected metrics for fixed bijections sorted according to the global cost obtained with all three datasets.}\label{agent-assignments:metric-heatmap}	
\vspace{-0.5cm}
\end{figure}

Figure~\ref{agent-assignments:metric-heatmap} illustrates the learning performance of the 39 selected metrics, listed in an ascending order\footnotemark[12]. Note that metrics such as \textsf{\footnotesize ASC-min-corr-pearson} or \textsf{\footnotesize ASC-min-dst1-coeff} have high performance in all three datasets and they are likely to capture fundamental structural characteristics influencing collective learning. The metric \textsf{\footnotesize DESC-min-value} reaches the $25^{th}$ percentile, while its ascending version reaches the $51^{st}$ percentile. This confirms the earlier intuition about the higher impact of agents with influential plan values when placed closer to the root. Note also the \textsf{\footnotesize avg-min-dst1-coeff} metric: it results in the highest performance of the $5^{th}$ percentile for agents sorted in descending order with a standard deviation of 0.004 among datasets indicating consistent behavior. In contrast, the ascending version of this metric results in one of the lowest performing solutions close to $73^{rd}$ percentile. Respectively, the \textsf{\footnotesize min-dst1-coeff} metric is on the $24^{th}$ and $76^{th}$ percentile for descending and ascending order. This polarization on sorting is also observed for \textsf{\footnotesize avg-min-dst3-coeff} and \textsf{\footnotesize min-dct3-coeff}.



Overall, averaging metrics do not usually perform well or they are around the mean percentile. Discrete sine and cosine transformations may result in alternating positive and negative values that average to zero causing information loss. Instead, metrics that rely on minimal or maximal values convey more information. The ascending order maximizes or minimizes performance and vice versa for the descending order. 

\begin{figure}[!htb]
\centering	
\subfloat[Synthetic dataset]{\includegraphics[width=1.0\columnwidth]{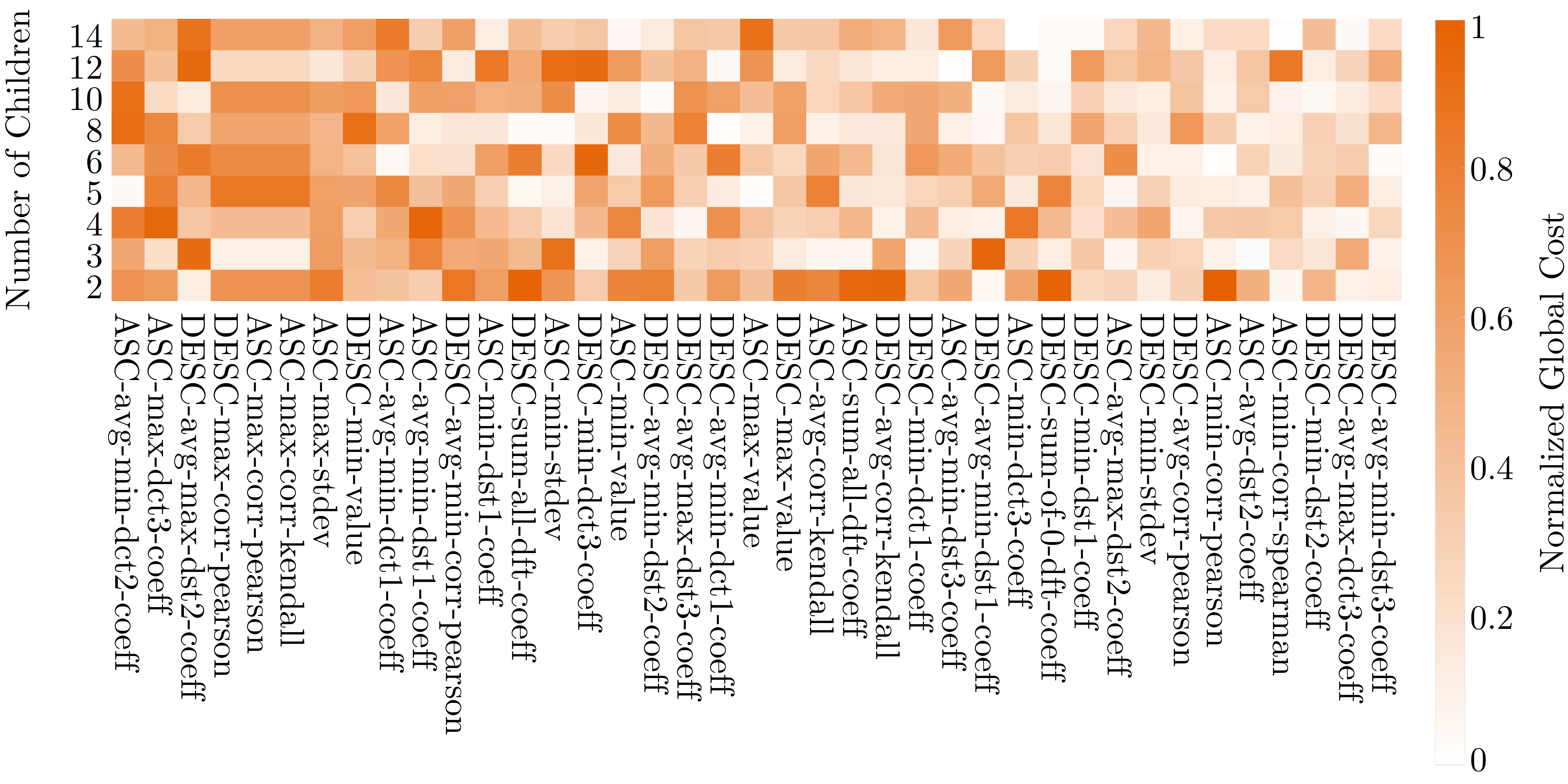}\label{agent-assignments:metric-heatmap:gaussian}}\\	
\subfloat[Energy dataset]{\includegraphics[width=1.0\columnwidth]{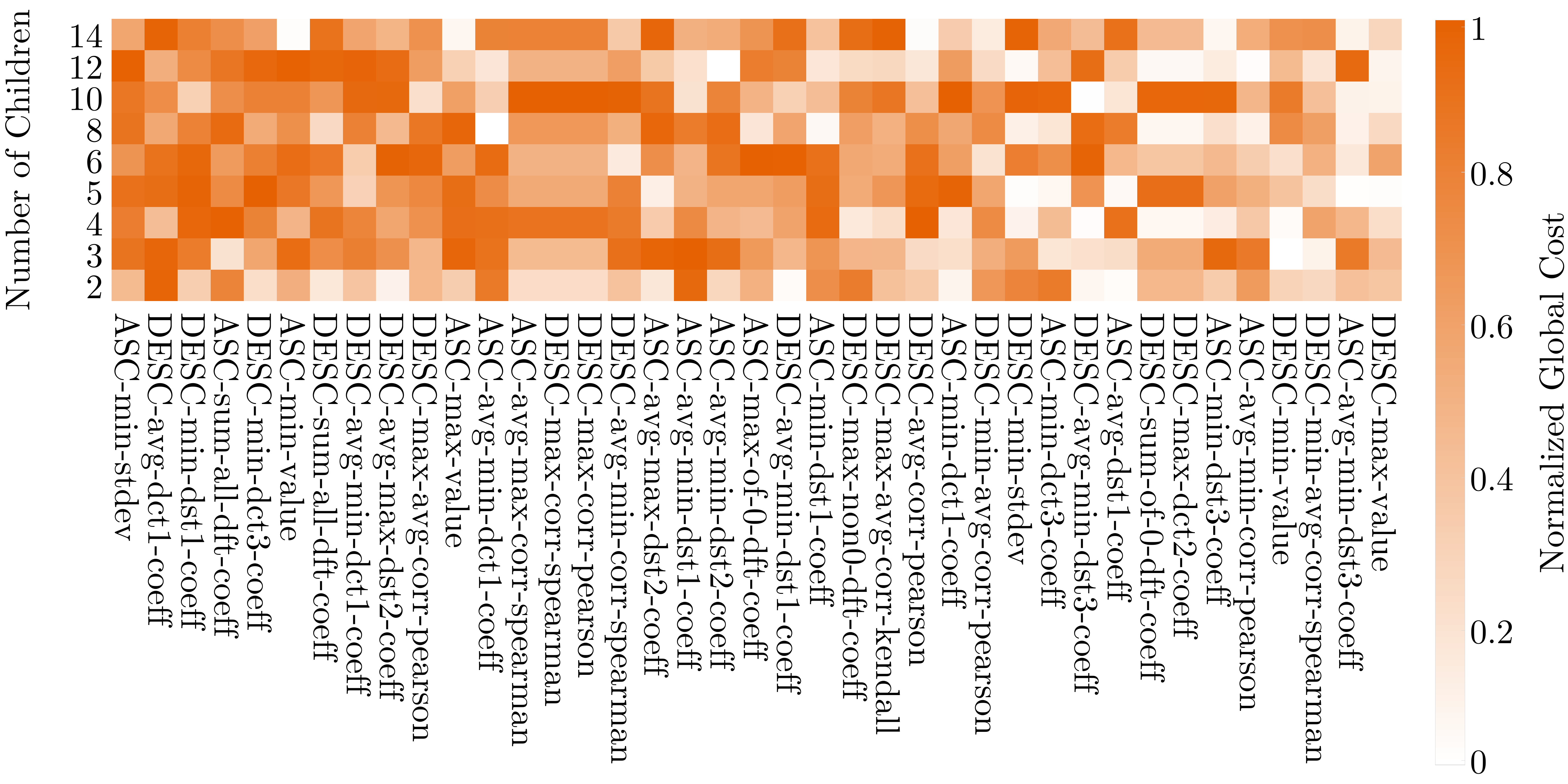}\label{agent-assignments:metric-heatmap:energy}}\\
\subfloat[Bicycle dataset]{\includegraphics[width=1.0\columnwidth]{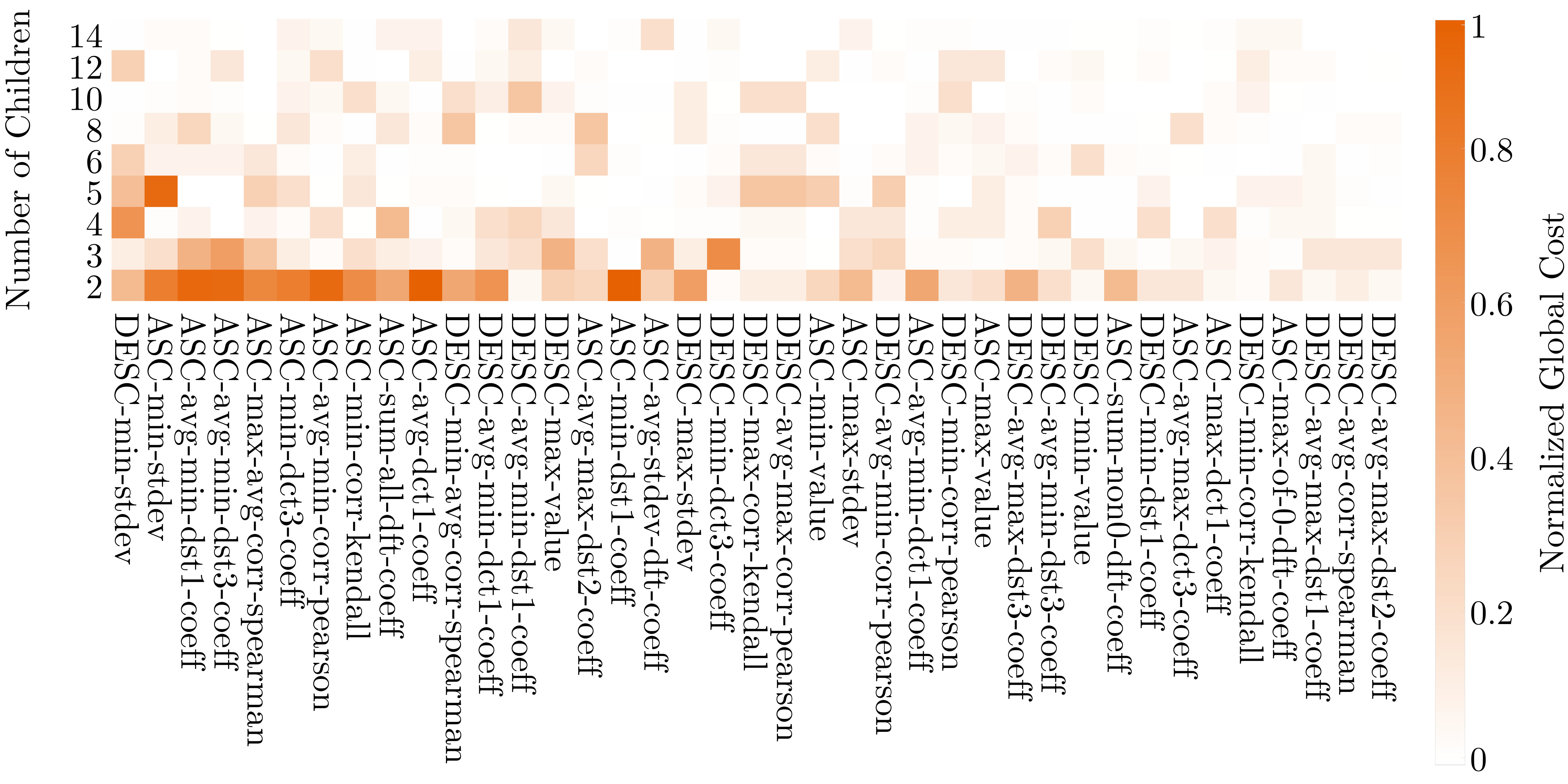}\label{agent-assignments:metric-heatmap:bicycle}}\\
\caption{Learning performance of selected metrics for fixed bijections under varying number of children and sorted according to the global cost obtained with all three datasets.}\label{agent-assignments:metric-heatmap-datasets} 
\vspace{-0.3cm}	
\end{figure}

Figure~\ref{agent-assignments:metric-heatmap-datasets} illustrates the optimality of fixed bijections under a varying number of children in the range $[2,14]$. Although structure does not show large influence on the synthetic and energy dataset, for the bicycle dataset with sparser plans~\cite{pournaras2018} the agents' positioning plays a more significant role. The following observations can be made for the bicycle dataset: (i) On average, the learning performance improves for all metrics by increasing the number of children per agent. (ii) The impact of the agents' positioning is significantly higher for a binary tree as a higher deviation is observed among the metrics compared to trees with more than 8 children per agent. 

\subsection{Learning performance with dynamic bijections}\label{subsec:dynamic-bijections}

\begin{figure}[!htb]
\vspace{-0.3cm}
\centering	
\subfloat[Synthetic dataset, $10^{th}$ percentile]{\includegraphics[width=0.327\columnwidth]{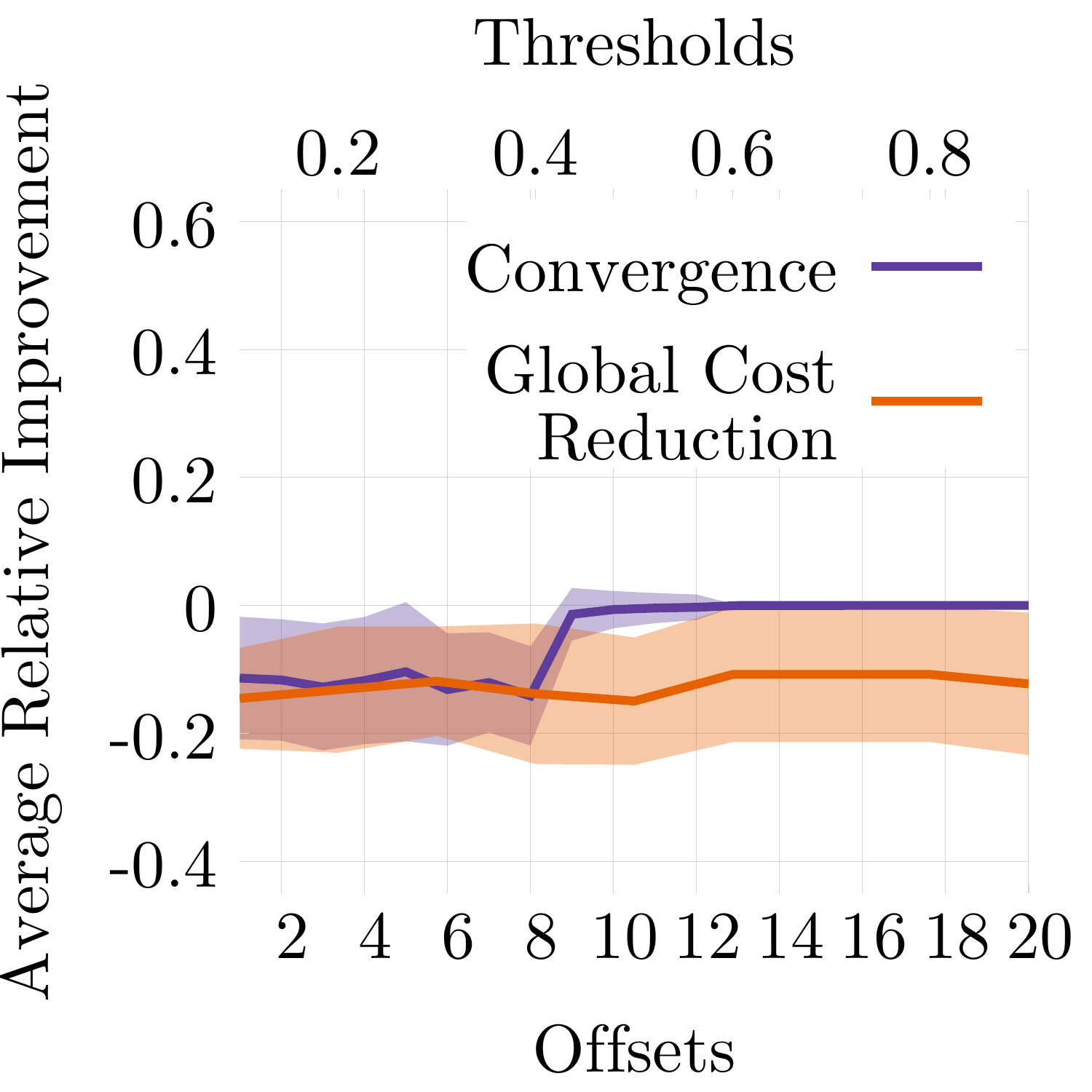}\label{comparison:improvement-10:gaussian}}\hfill
\subfloat[Synthetic dataset, $50^{th}$ percentile]{\includegraphics[width=0.327\columnwidth]{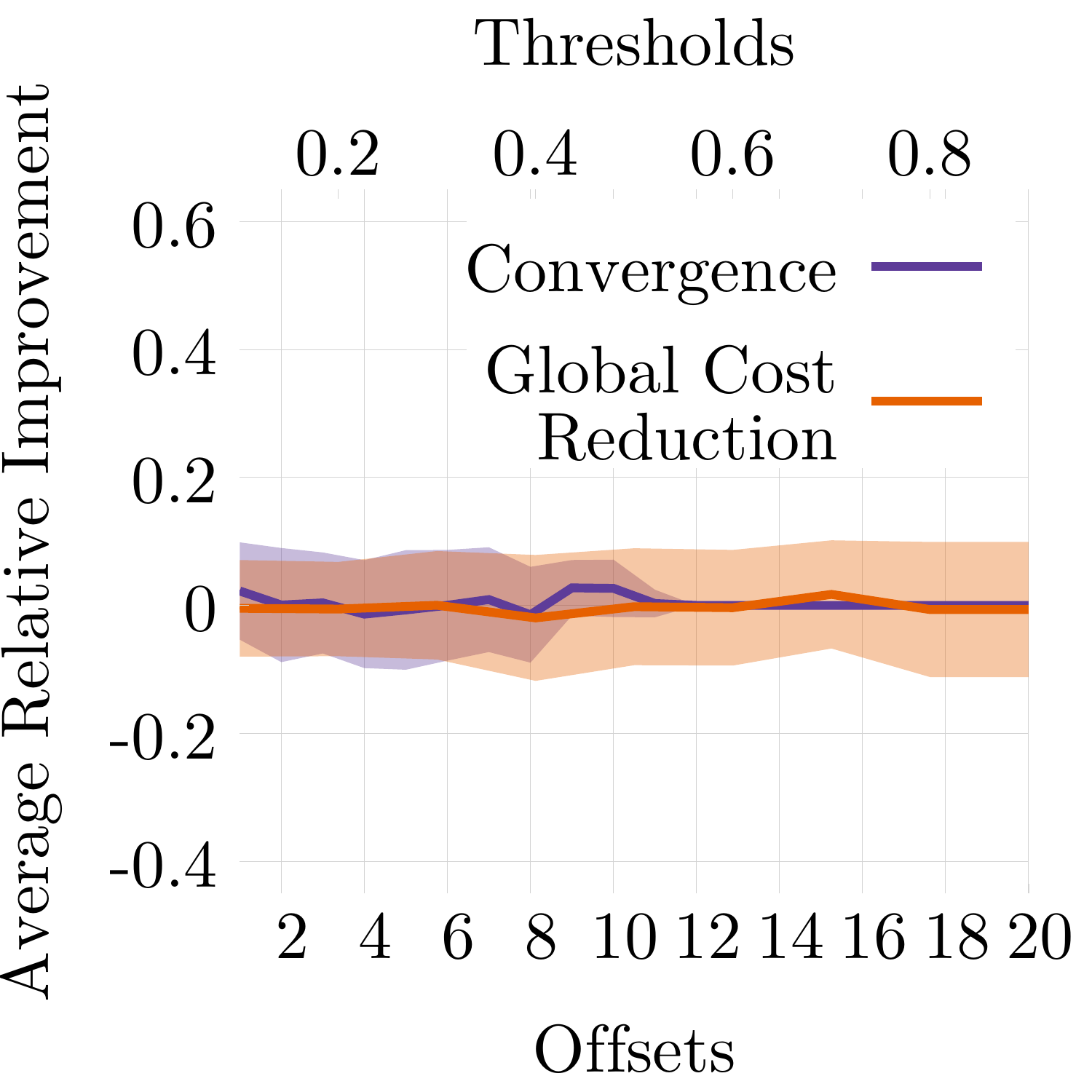}\label{comparison:improvement-50:gaussian}}\hfill
\subfloat[Synthetic dataset, $90^{th}$ percentile]{\includegraphics[width=0.327\columnwidth]{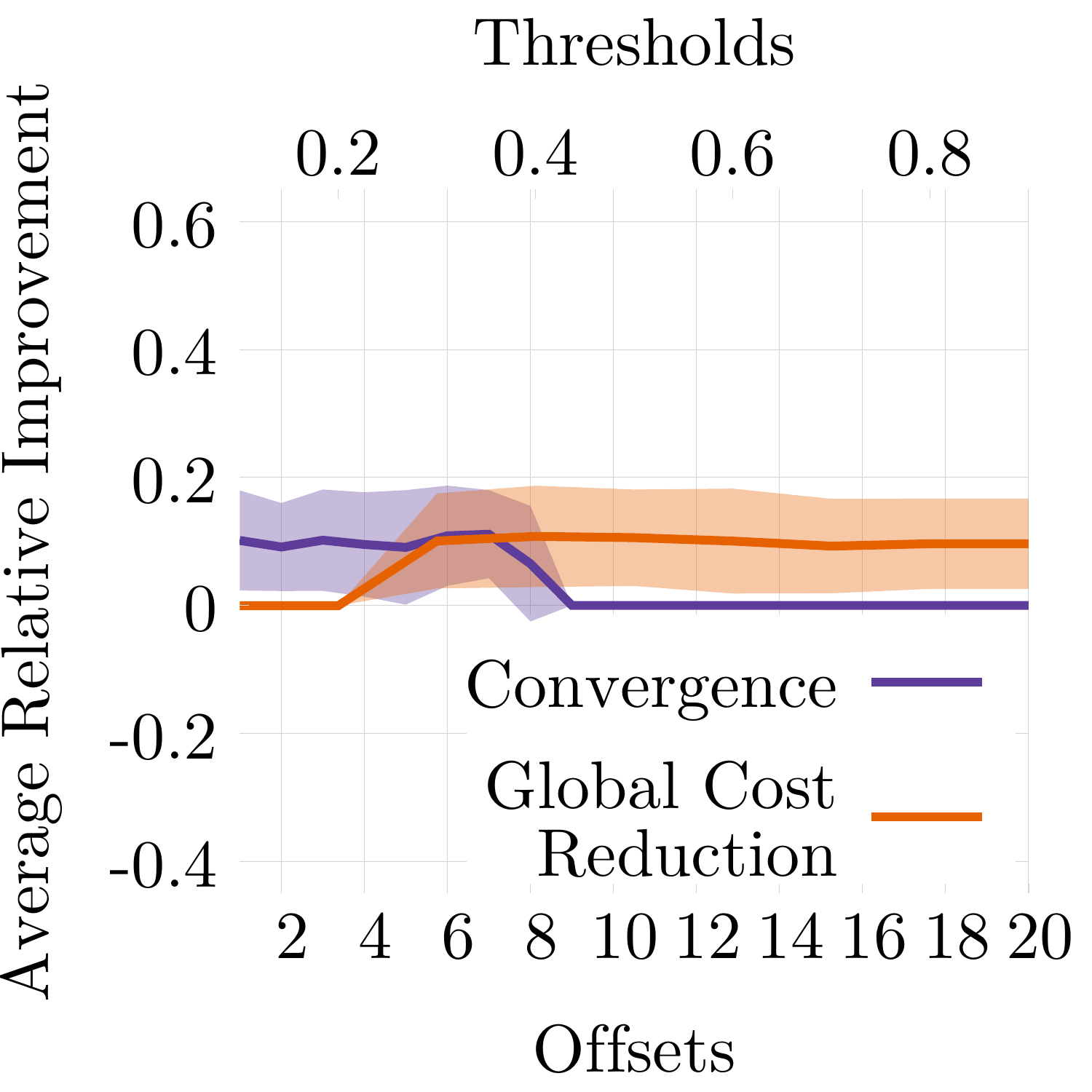}\label{comparison:improvement-90:gaussian}}\\
\subfloat[Energy dataset, $10^{th}$ percentile]{\includegraphics[width=0.327\columnwidth]{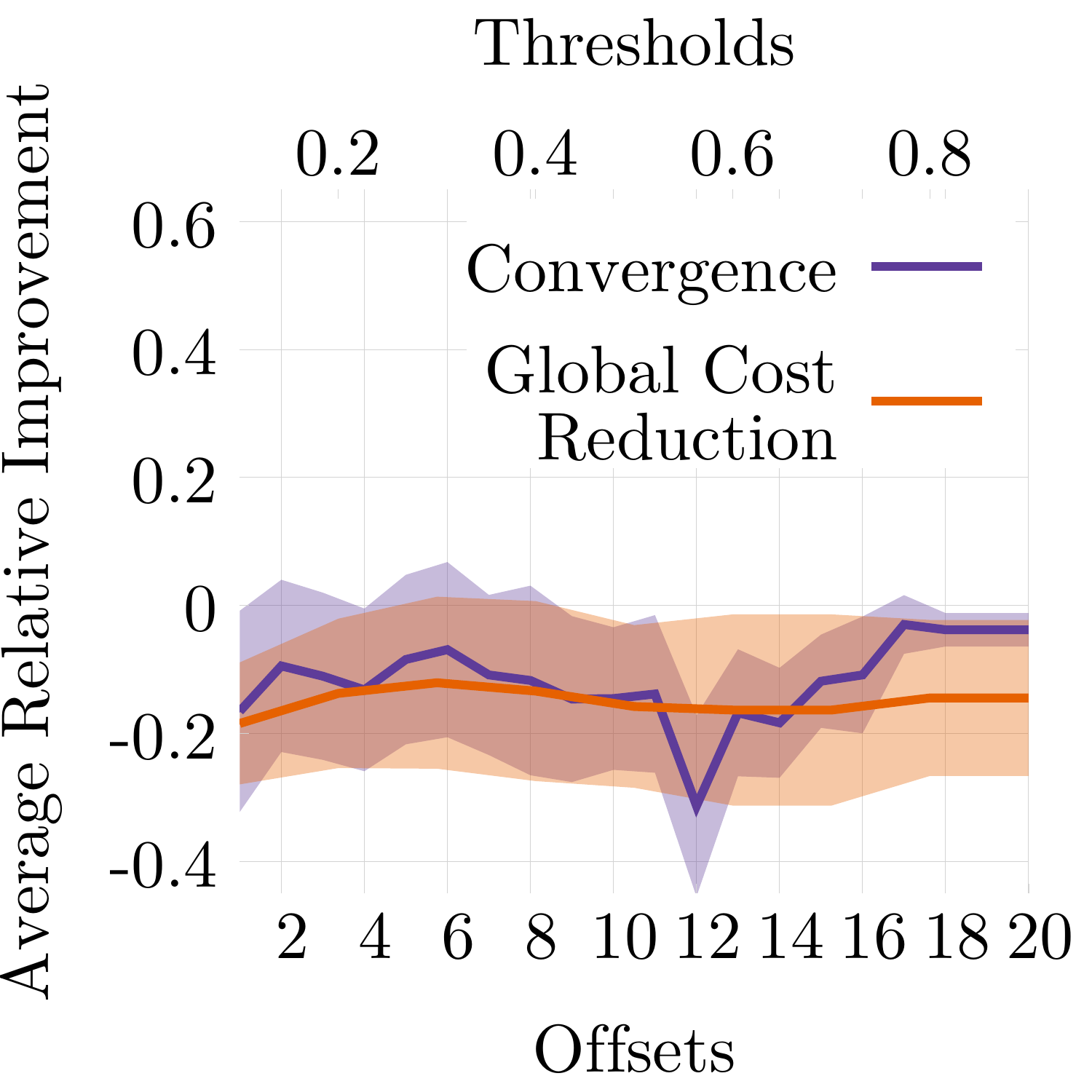}\label{comparison:improvement-10:energy}}\hfill
\subfloat[Energy dataset, $50^{th}$ percentile]{\includegraphics[width=0.327\columnwidth]{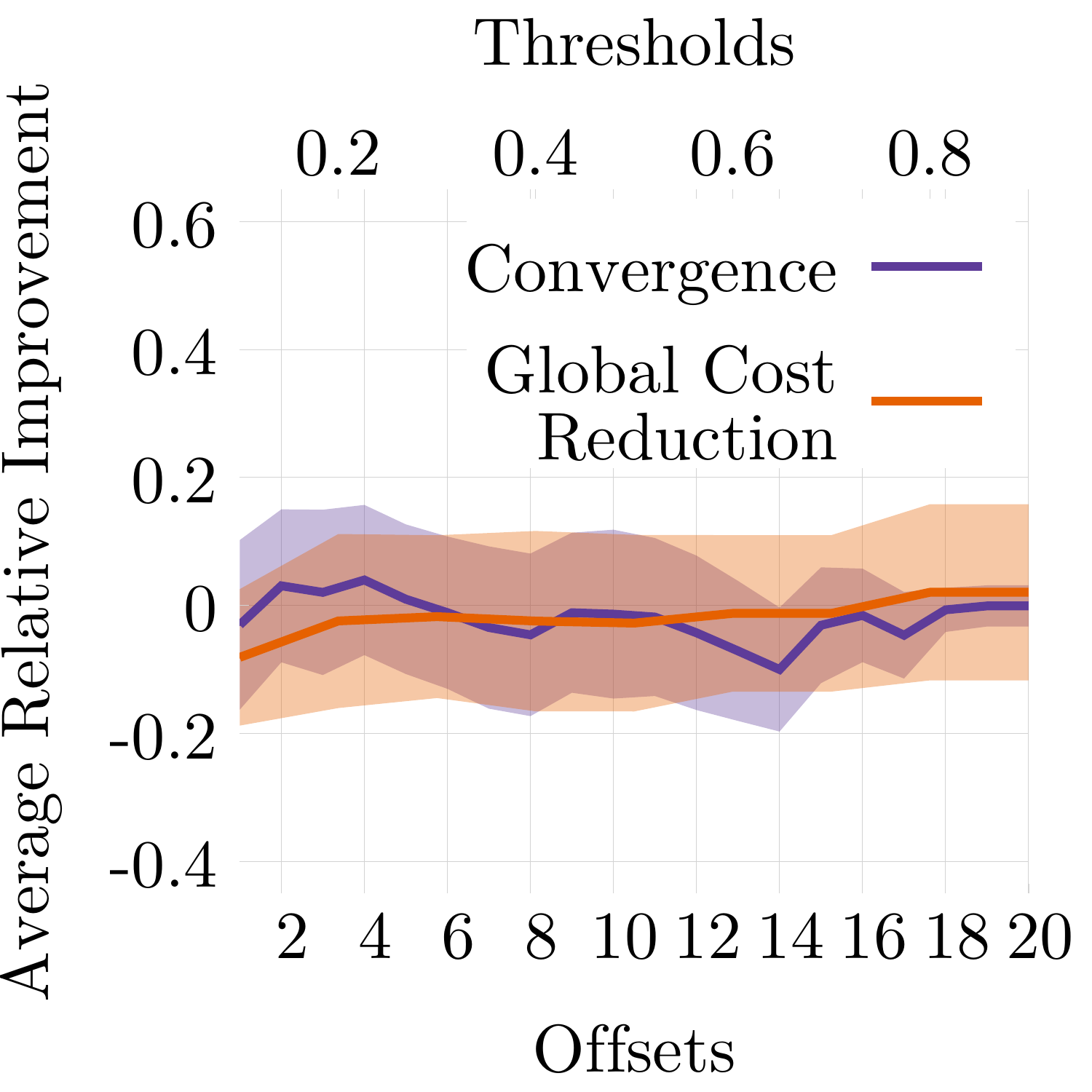}\label{comparison:improvement-50:energy}}\hfill
\subfloat[Energy dataset, $90^{th}$ percentile]{\includegraphics[width=0.327\columnwidth]{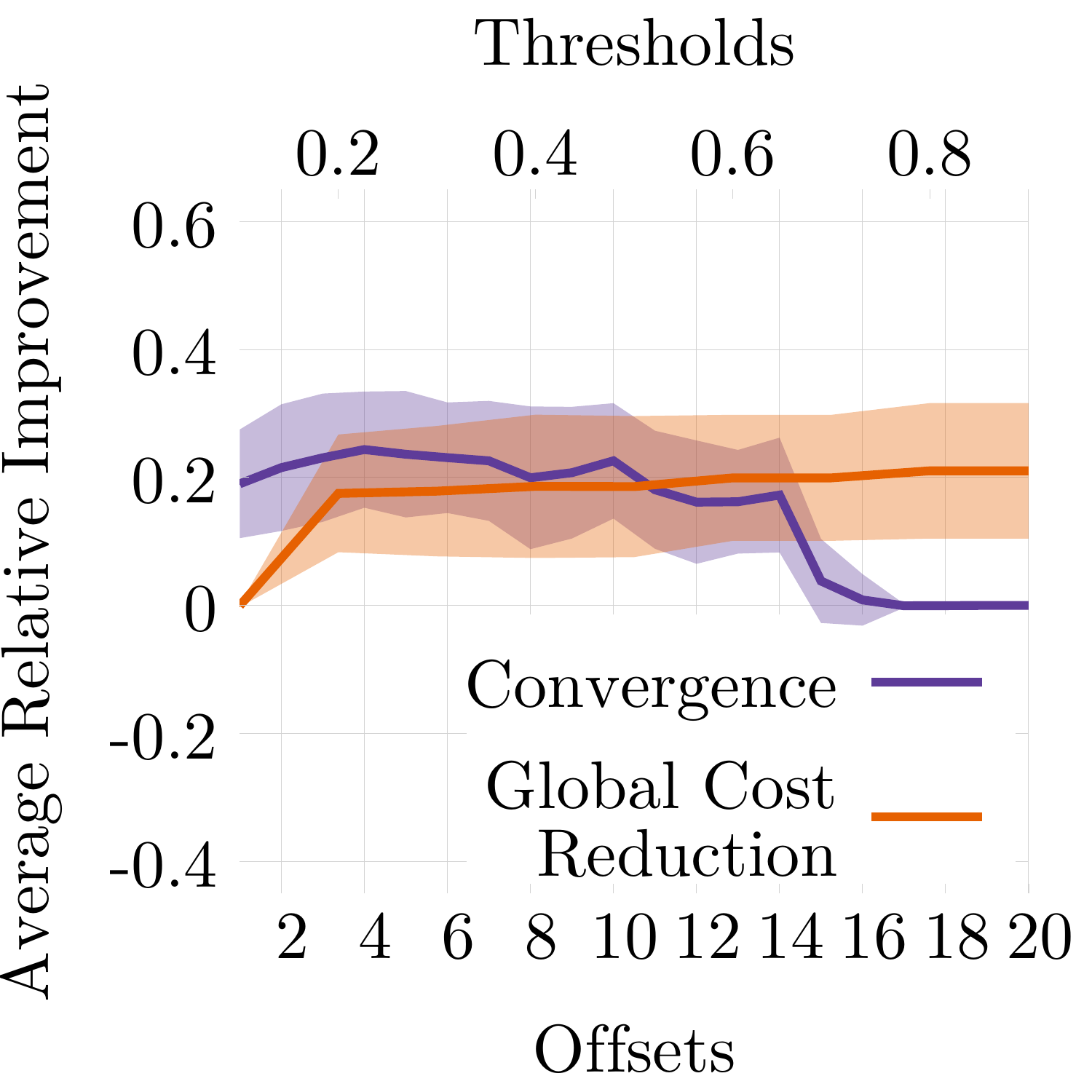}\label{comparison:improvement-90:energy}}\\
\subfloat[Bicycle dataset, $10^{th}$ percentile]{\includegraphics[width=0.327\columnwidth]{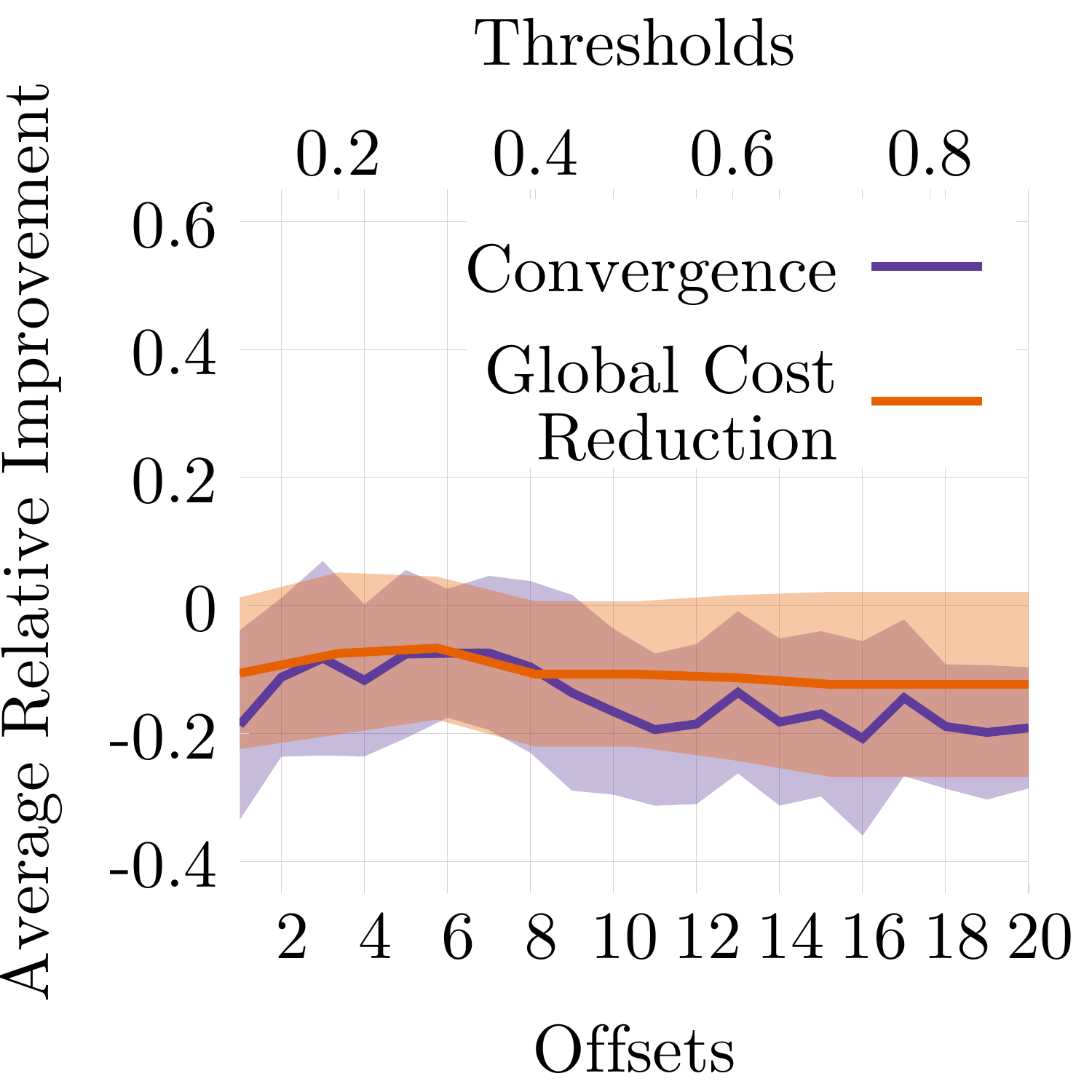}\label{comparison:improvement-10:bicycle}}\hfill
\subfloat[Bicycle dataset, $50^{th}$ percentile]{\includegraphics[width=0.327\columnwidth]{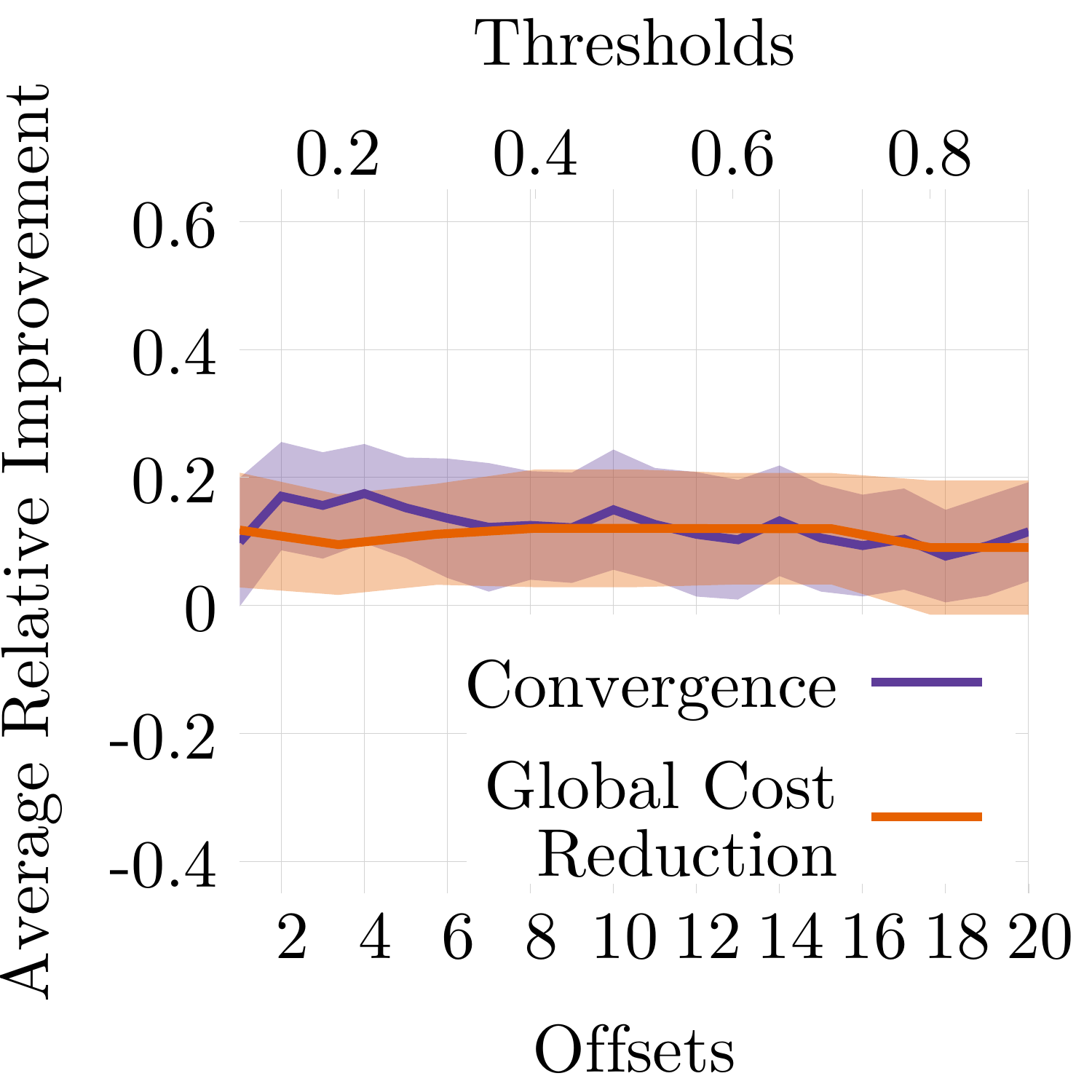}\label{comparison:improvement-50:bicycle}}\hfill
\subfloat[Bicycle dataset, $90^{th}$ percentile]{\includegraphics[width=0.327\columnwidth]{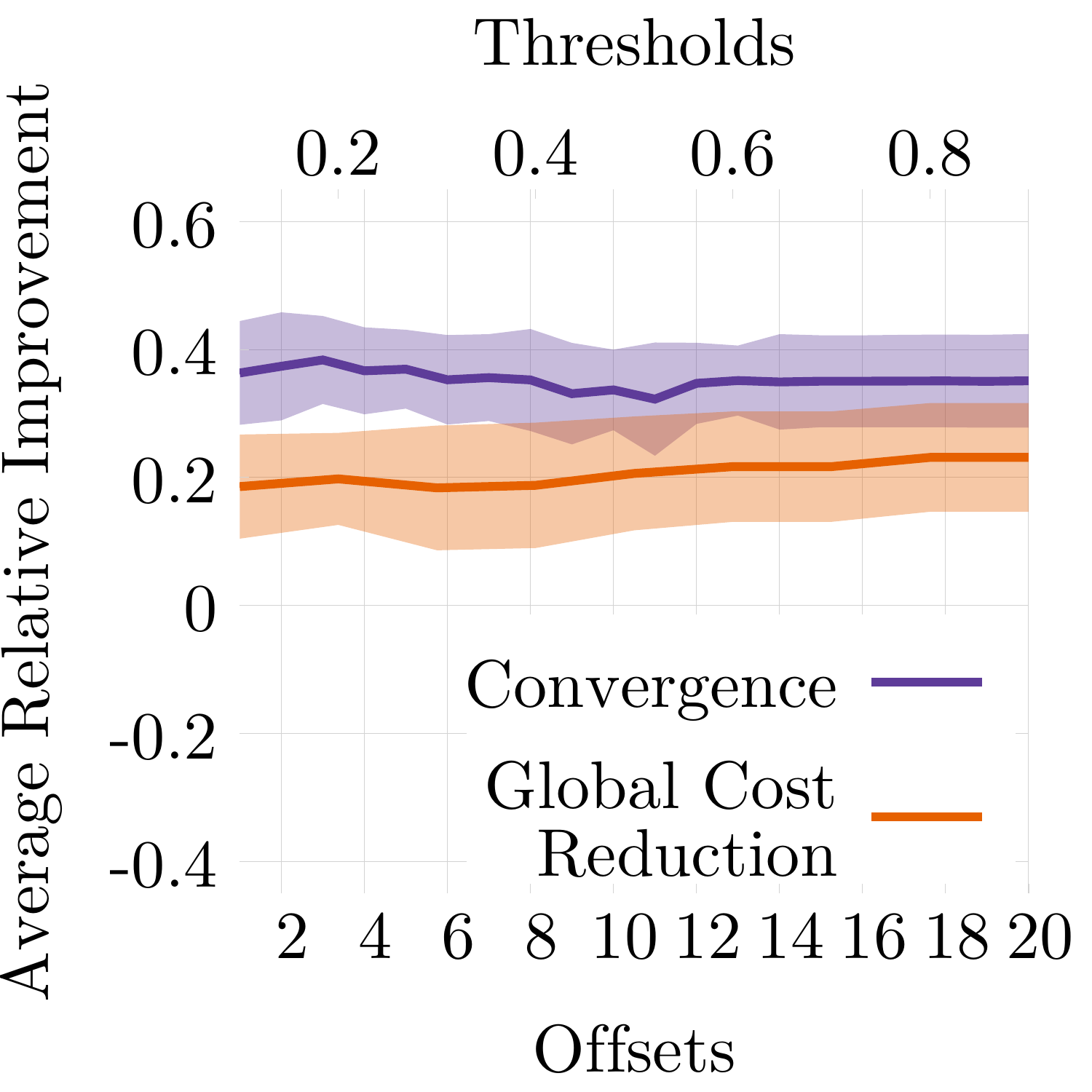}\label{comparison:improvement-90:bicycle}}\\
\caption{Learning performance of the two structural self-adaptation strategies for different datasets, baseline percentiles, memory offsets and thresholds.}\label{comparison:improvement} 	
\end{figure}

Figure~\ref{comparison:improvement} compares the two online strategies for different datasets, baseline percentiles, memory offsets and thresholds. The optimality of the initial tree structure significantly influences learning performance. Starting from the $10^{th}$ percentile, self-adaptations are likely to decrease performance, nevertheless, to a low extent and there are cases in which performance improves as indicated by the area representing the standard deviation. In contrast, the $90^{th}$ percentile results in significantly lower global cost. The $50^{th}$ percentile shows a low performance improvement, mainly for the bicycle dataset. Note though that in case improved solutions are memorized or highly performing fixed bijections are applied, as shown in Section~\ref{subsec:fixed-bijections}, learning performance can further improve.

\begin{figure}[!htb]
\vspace{-0.35cm}
\centering	
\subfloat[Synthetic dataset, convergence criterion with long-term memory]{\includegraphics[width=0.327\columnwidth]{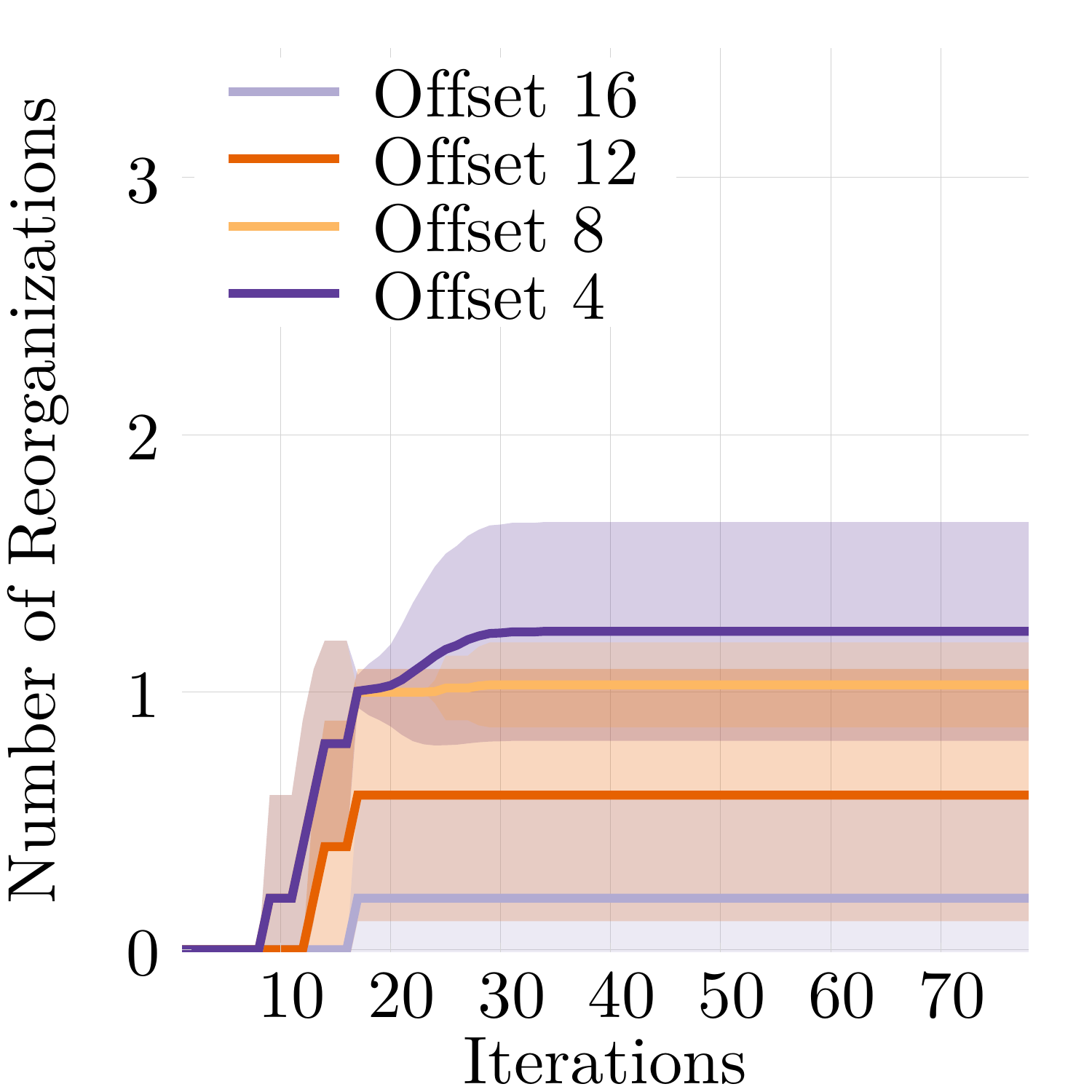}\label{agent-assignments:convergence:gaussian}}\hfill
\subfloat[Energy dataset, convergence criterion with long-term memory]{\includegraphics[width=0.327\columnwidth]{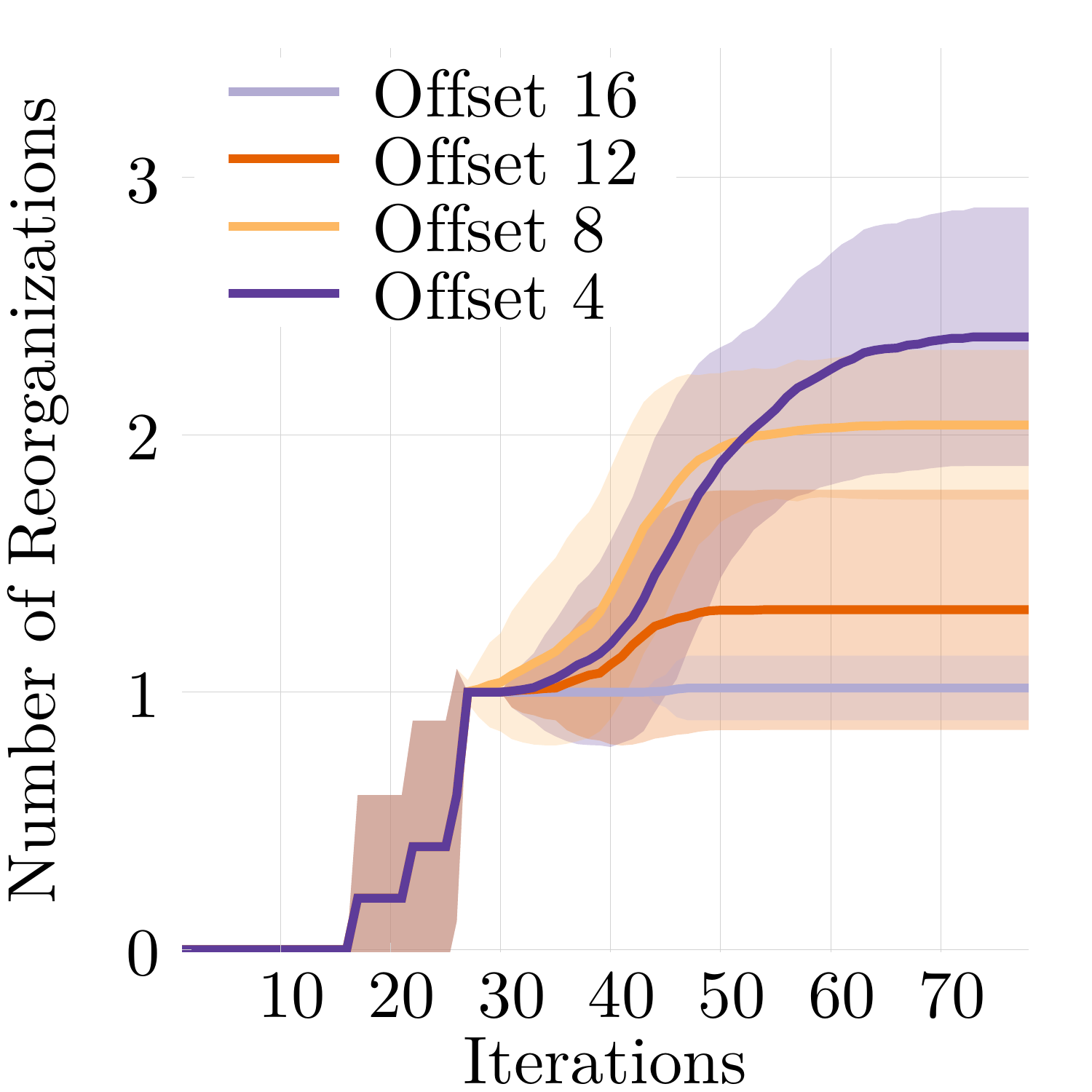}\label{agent-assignments:convergence:energy}}\hfill
\subfloat[Bicycle dataset, convergence criterion with long-term memory]{\includegraphics[width=0.327\columnwidth]{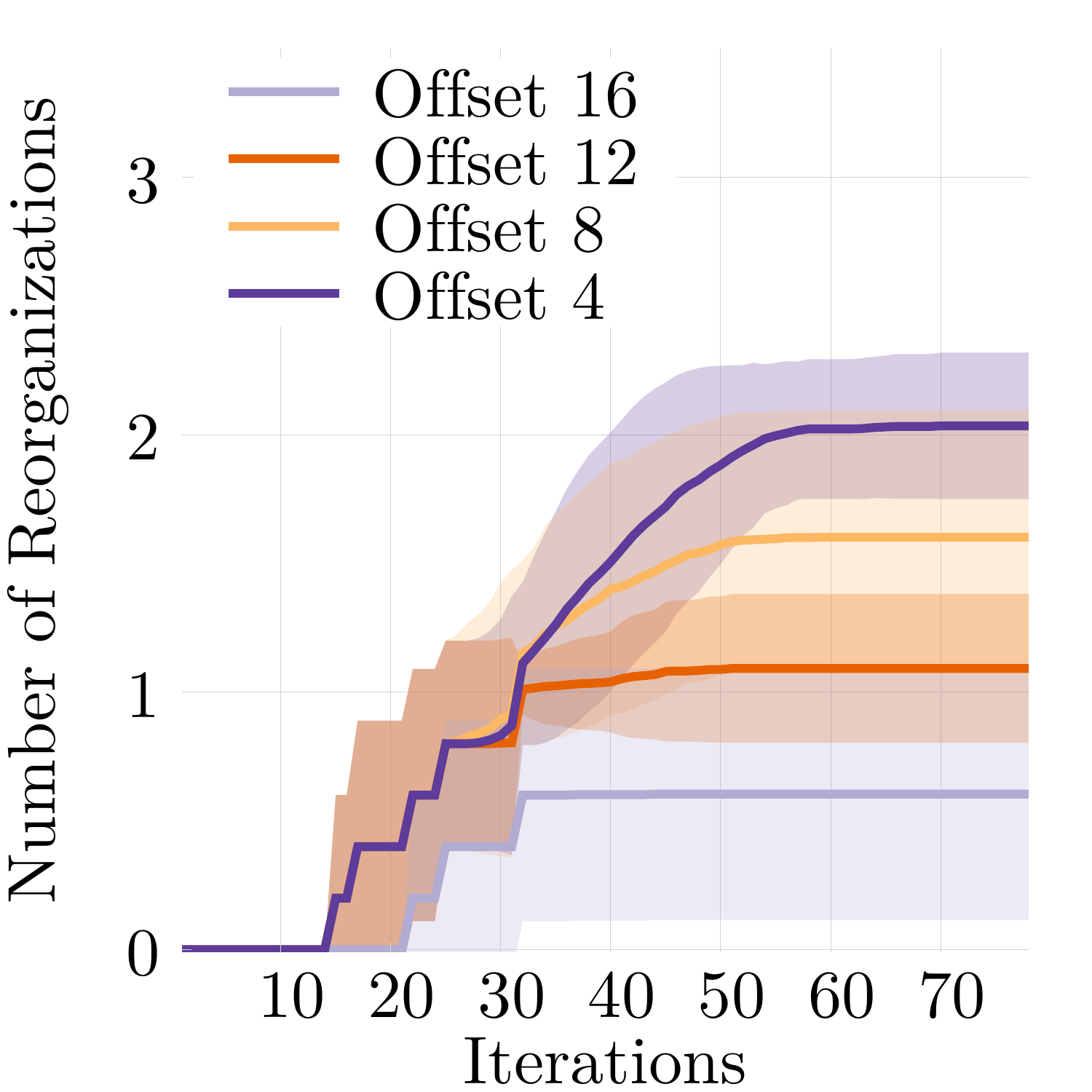}\label{agent-assignments:convergence:bicycle}}\hfill\\
\subfloat[Synthetic dataset, global cost reduction criterion with short-term memory]{\includegraphics[width=0.327\columnwidth]{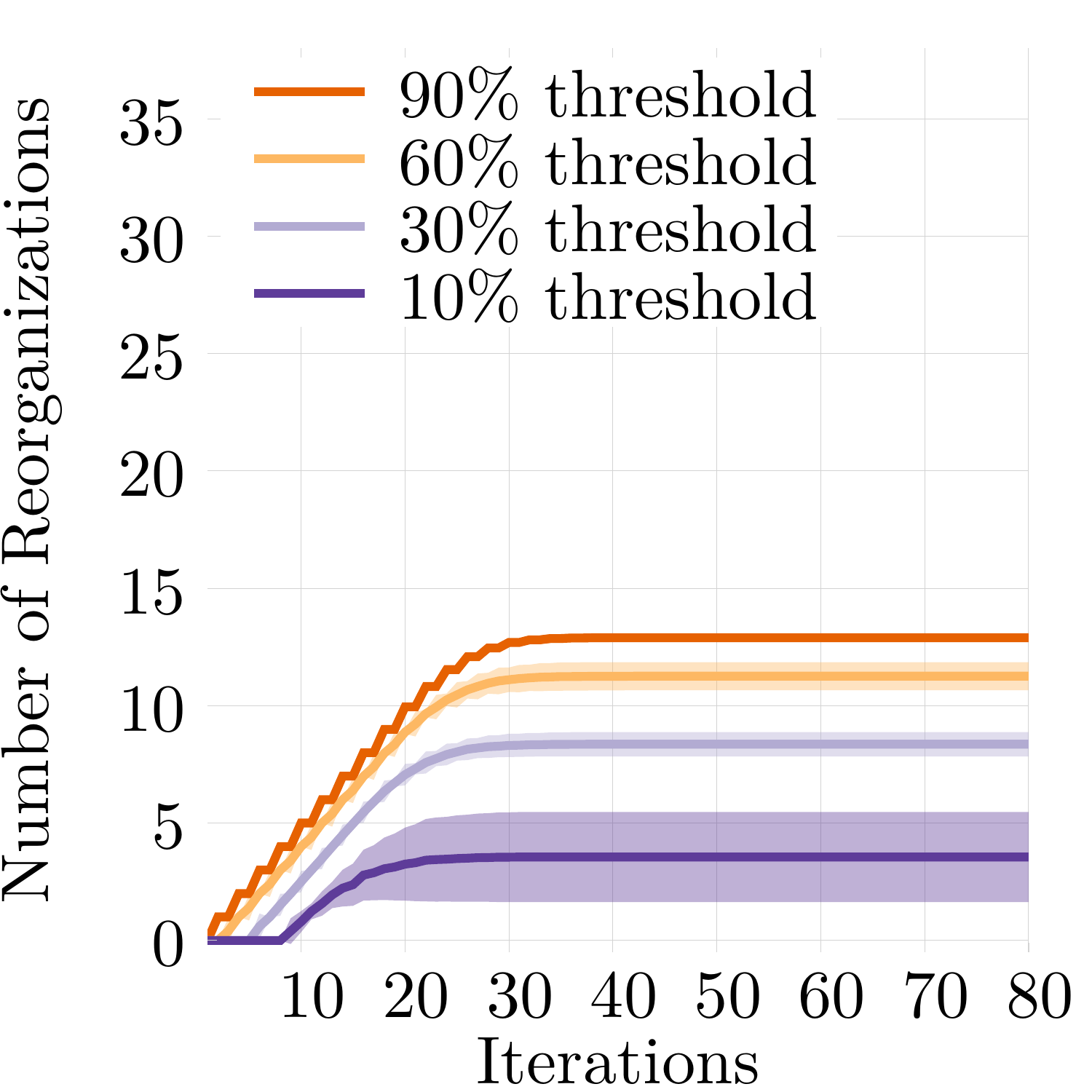}\label{agent-assignments:global-cost:gaussian}}\hfill
\subfloat[Energy dataset, global cost reduction criterion with short-term memory]{\includegraphics[width=0.327\columnwidth]{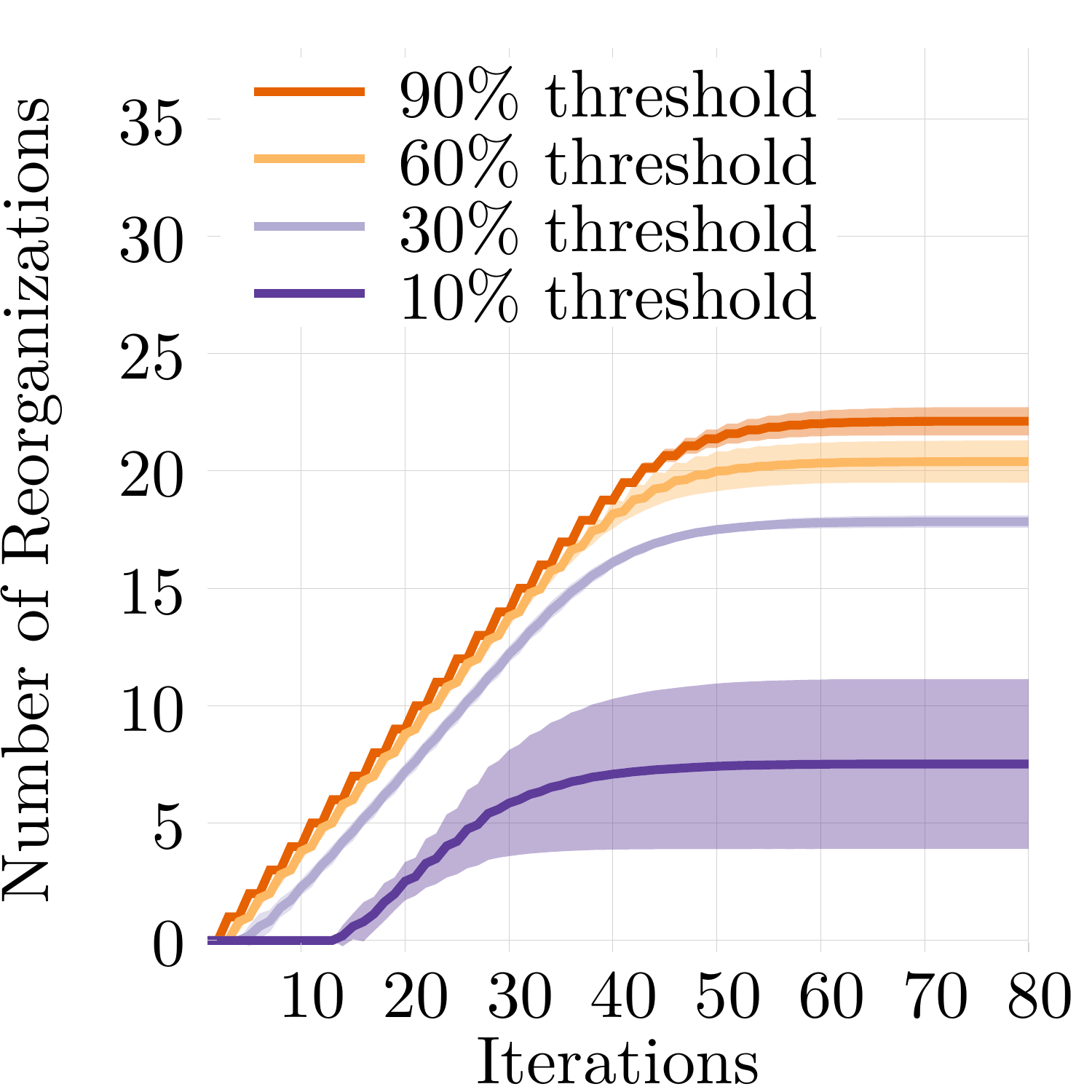}\label{agent-assignments:global-cost:energy}}\hfill
\subfloat[Bicycle dataset, global cost reduction criterion with short-term memory]{\includegraphics[width=0.327\columnwidth]{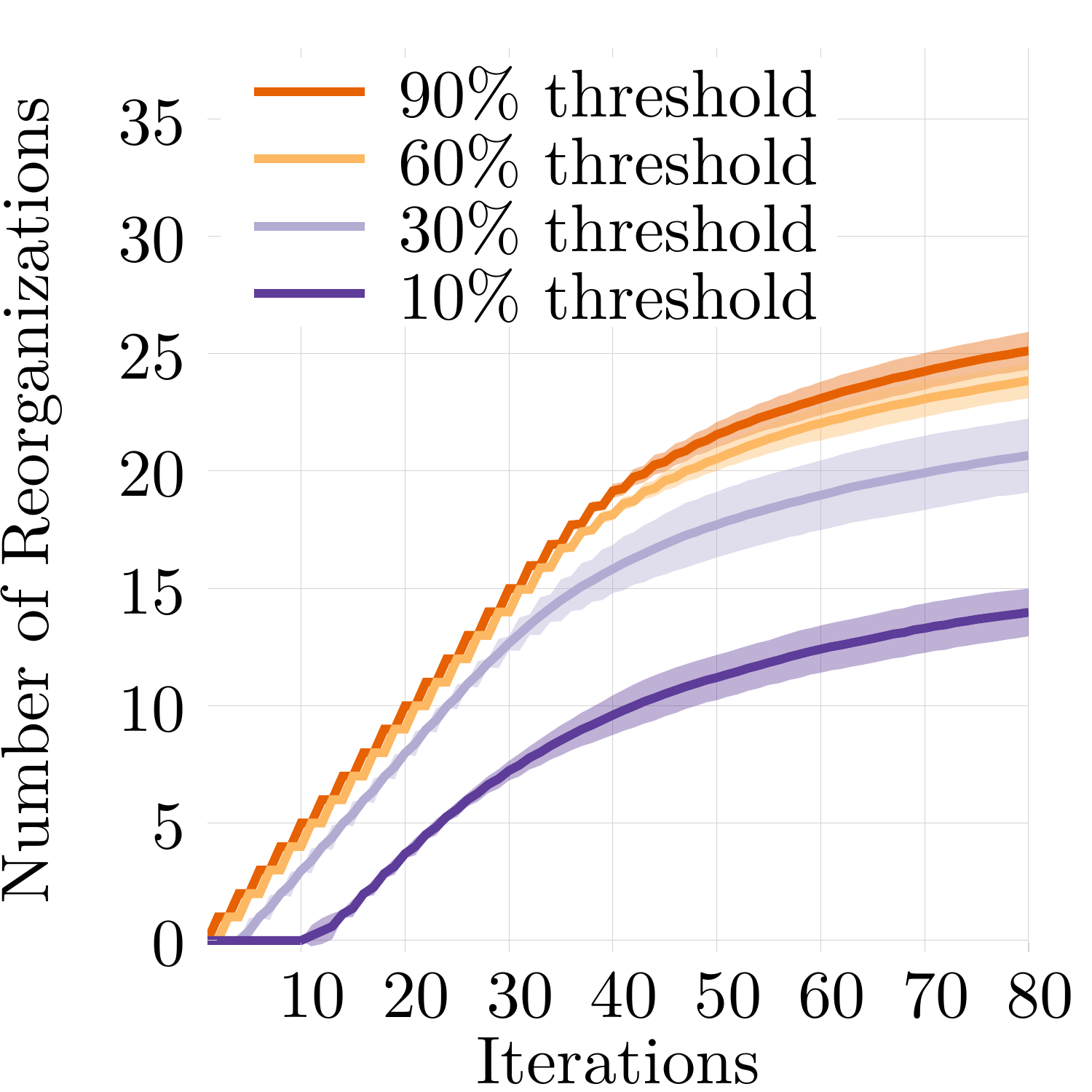}\label{agent-assignments:global-cost:bicycle}}\hfill
\caption{Cumulative number of self-adaptations over the learning runtime for the different strategies, memory offsets, thresholds and datasets.}\label{agent-assignments:num-reorganziations} 	
\vspace{-0.5cm}
\end{figure}

The following observations can be made for the two strategies: (i) On average, the relative improvement of the convergence criterion with long-term memory is $-0.3\%$, $0.33\%$ and $11.01\%$ for the bicycle, energy and synthetic datasets respectively. For the global cost reduction criterion with short-term memory the respective numbers are $-1.7$\%, $0.13\%$ and $7.03\%$. (ii) High memory offsets favor initialization with a highly performing tree structure ($10^{th}$ percentile), while low memory offsets favor initialization with a low performing tree ($90^{th}$ percentile). The likelihood of improving further a high-performing learning structure is lower unless solutions close to convergence are memorized. On the other hand, low-performing learning structures benefit from exploration with low memory offsets. (ii) The convergence criterion with long-term memory and low offsets maximize performance optimality across all baseline percentiles and datasets. (iii) Learning can be trapped into the first suboptimal solution for offsets higher than 10 using the convergence criterion with long-term memory as well as for threshold values lower than 0.2 using the global cost reduction with short-term memory, 

Figure~\ref{agent-assignments:num-reorganziations} illustrates the cumulative number of self-adaptations during the learning runtime for each of the two strategies and for different offsets, thresholds and datasets. Results are averaged out over all benchmark percentiles. 

The convergence criterion with long-term memory results in significantly lower structural self-adaptations. I-EPOS requires very few iterations, 10 or 15 for instance~\cite{pournaras2018}, to converge. Dividing the number of iterations with the offset iteration indicates the number of structural self-adaptation to expect. The higher the offset, the higher the optimality of the memorized solution and as a result the higher the likelihood of early termination at the next learning phase. 

The global cost reduction criterion with short-term memory shows a higher number of self-adaptations, especially for the energy and bicycle dataset. The termination criterion measuring the slope of the global cost prevents early termination of I-EPOS by triggering self-adaptation early before convergence. This is especially the case for higher thresholds. Self-adaptations are triggered throughout the learning runtime for the bicycle dataset, while termination is observed around the $25^{th}$ and $40^{th}$ iteration for the synthetic and energy dataset. 

\section{Conclusion and Future Work}\label{sec:conclusion}

This paper concludes that structure has a foundational role for the cost-effectiveness of decentralized pervasive intelligence as confirmed by the following: (i) Deterministic meta-feature criteria with which fixed structural self-adaptations are performed influence learning performance. (ii) Online structural self-adaptations can improve learning performance and prevent suboptimal trapped solutions, especially under low-performing initialization. A large-scale benchmark dataset for optimality evaluation is made openly available. It relies on real-world datasets of residential power demand, bike sharing and charging of electric vehicles. Millions of structural self-adaptation in networks of thousands of agents exchanging billion of learning messages provide representative performance profiles as shown with a comparison of parametric vs. non-parametric density estimations of the solutions space. 

The construction of smarter online structural self-adaptations that consider the communication and computational cost of the agents' repositioning is part of future work. Studying other communication structures besides trees can provide further insights and fundamental understanding. 

\vspace{-0.15cm}

\bibliography{bibliography} 
\bibliographystyle{IEEEtran}

\end{document}